\documentclass[aps,prb,twocolumn,superscriptaddress,showpacs]{revtex4-1}

\usepackage{graphicx}
\usepackage{dcolumn}
 
\usepackage{amsmath}

\graphicspath{ {figs/} }

\begin{document}


\title{Magnetic Properties of GeNi$_{2}$O$_{4}$ Under High Pressure and Magnetic Dilution}


\author{J.T. Korobanik}
\email[Jory Korobanik] {jory.korobanik@brocku.ca}
\affiliation{Dept. of Physics, Brock University. 500 Glenridge Ave, St Catharines, ON, L2S 3A1, Canada}
\author{K. Caslin}
\affiliation{Dept. of Physics, Brock University. 500 Glenridge Ave, St Catharines, ON, L2S 3A1, Canada}
\affiliation{Max-Planck-Institute for Solid State Research, Heisenbergstrasse 1, D-70569 Stuttgart, Germany}
\author{P. Reuvekamp}
\affiliation{Max-Planck-Institute for Solid State Research, Heisenbergstrasse 1, D-70569 Stuttgart, Germany}
\author{F.S. Razavi}
\affiliation{Dept. of Physics, Brock University. 500 Glenridge Ave, St Catharines, ON, L2S 3A1, Canada}


\date{\today}

\begin{abstract}
The effects of magnetic dilution and applied pressure on frustrated spinel Ni$_{2-x}$Mg$_{x}$GeO$_{4}$ (0 $\leq x \leq 1.0$) are analyzed using specific heat, AC and DC magnetization and x-ray diffraction. The parent compound has two closely spaced antiferromagnetic transitions $T_{\text{N1}}=12.0$ K (kagom\'{e} planes) and $T_{\text{N2}}=11.4$ K (triangular planes). In the dilution range tested the low temperature magnetic state takes three forms: antiferromagnetic ($0 \leq x \leq 0.05$), ill-defined ($x=0.10$ and 0.15), and spin glass ($0.30 \leq x \leq 1.0$). The AFM region shows an extreme vulnerability to dilution with a percolation threshold of $p_{\text{c1}}=0.74 \pm 0.04$ and $p_{\text{c2}}=0.65 \pm 0.05$ for the kagom\'{e} and triangular planes respectively, which are much larger than expected for 3D systems. We suggest that this behavior is due to coupling between the kagom\'{e} and triangular spins forming a 'network of networks' (NON). Thermal expansion data on parent NGO indicates a field dependent lattice contraction during ordering events. Furthermore, there is a transition from contraction to expansion in an applied field of 6 T in the kagom\'{e} planes. For dilution levels $x \geq 0.30$, the system becomes a spin glass with canonical behavior. Specific heat results suggest that the triangular spins become disordered with increasing Mg$^{2+}$ substitution followed by the onset of glassiness in the kagom\'{e} planes. Furthermore, there appears to be a dilution driven shift from 3 dimensional to 2 dimensional behavior as the low temperature magnetic heat capacity scales as $T^{2}$ in the spin glass state.

\end{abstract}

\pacs{75.50.Lk, 75.30.Kz, 75.40.-s}

\maketitle


\section{Introduction}

Geometrically frustrated materials have garnered much research attention due to the multitude of complex low temperature states they exhibit.\cite{Wiebe2015, Balents2010,Ji2009, Castelnovo2008, Fischer1993} Magnetic analogues of structurally disordered systems such as spin ice\cite{Harris1997} and spin glass~\cite{Binder1986} have been extensively studied.\cite{Diep2013} The essence of magnetic frustration is the inability to simultaneously satisfy magnetic interactions. When this arises from the physical layout of magnetic ions, it is term geometric frustration.\cite{Kang2014} This competition between magnetic interactions can preclude the formation of a distinct magnetic ground state.\cite{Ramirez1994} The low temperature state is often determined by small perturbations such as dipole-dipole interactions, structural changes and impurities which act to lift the degeneracy and relieve frustration.\cite{Greedan2001} Typically, this results in a $T_{\text{c}}$ at lower temperatures due to the effective quenching of exchange. Geometric frustration often occurs in triangular based structures under antiferromagnetic (AFM) nearest neighbor exchange. Examples of these are edge-sharing triangular and corner-sharing triangular (kagom\'{e}) lattices. Three dimensional realizations of these often occur in materials with pyrochlore and spinel structures.~\cite{Gardner2010}

\begin{figure}[hf]
 \includegraphics[width=0.45\textwidth]{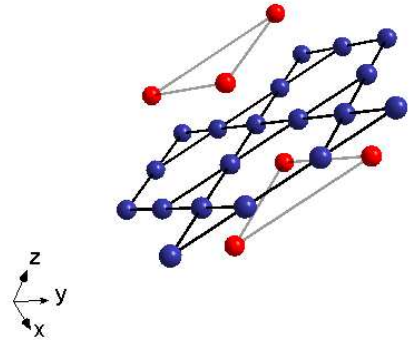}
 \caption{[color online] Simplified illustration of B-site sub-lattice in GeNi$_{2}$O$_{4}$ spinel. All spheres are Ni$^{2+}$ ions with blue representing those in kagom\'{e} layers and red representing those in triangular layers. \label{fig:lattice}}
 \end{figure}

GeNi$_{2}$O$_{4}$ (NGO) crystallizes in olivine or spinel structure types depending on growth conditions.\cite{Navrotsky1976} Materials that crystallize with spinel structure (AB$_{2}$O$_{4}$) are composed of B sites with  octahedral coordination and A sites with tetrahedral coordination. One can view this structure as two inter-penetrating diamond and pyrochlore sub-lattices that are associated with either the A-site (diamond) or B-site (pyrochlore). When viewing the B sub-lattice along the [111] direction, one observes alternate stacking of kagom\'{e} and triangular planes which is highlighted in Fig. \ref{fig:lattice}. Geometric magnetic frustration can result from an AFM exchange between B site ions (B=Mn$^{3+}$, Cr$^{3+}$, V$^{3+}$, Fe$^{3+}$).\cite{Lacroix2011} The parameter\cite{Ramirez1994} $f=| \theta_{\text{cw}}|$/$T_{\text{c}}$, which is a measure of N\'{e}el suppression, is often used to characterize the strength of frustration. For example, the spinel ZnCr$_{2}$O$_{4}$ has a very high value f=31.2 and has been shown to order after a lattice distortion from cubic to tetragonal.\cite{Lee2002} Neutron diffraction studies have indicated the formation of a complex co-planar spin state at low temperature.\cite{Ji2009} Regarding Germanates, several compounds form with spinel structures such as GeCo$_{2}$O$_{4}$\cite{Lashley2008} and, under high pressure conditions,  GeCu$_{2}$O$_{4}$\cite{Yamada2000}.  

Interest in GeNi$_{2}$O$_{4}$ has grown due to the existence of two closely spaced magnetic transitions at $T_{\text{N1}}=12.0$ K and $T_{\text{N2}}=11.4$ K.\cite{Crawford2003} Neutron diffraction studies\cite{Matsuda2008} on single crystals indicate that these ordering phenomena are due to the alignment of spins within the stacked kagom\'{e} and triangular planes. Specifically, at $T_{\text{N1}}$ the spins within the kagom\'{e} planes align ferromagnetically with AFM alignment between successive like-planes. At $T_{\text{N2}}$ the triangular planes arrange themselves in the same fashion. To describe this ordering behaviour a large exchange network with a dominant fourth nearest-neighbour J$_{4}$ term is required.\cite{Matsuda2008} The degeneracy of 3d orbitals for Ni$^{2+}$ ions ($^{3}F_{4}$) in octahedral crystal field is lifted with a ground state of S=1. The triplet t$_{2g}$ levels are further split by spin-orbit coupling but this contribution is very small and can be ignored.\cite{Lashley2008,Ashcroft1976} Specific heat data have shown that only approximately 60$\%$ of the expected magnetic entropy is recovered. In addition, the entropy change associated with each ordering event is nearly the same which is unusual as three times as many spins reside on the kagom\'{e} compared to the triangular planes. 

Magnetic dilution is an important technique in determining information on the several parameters about the exchange interactions in magnetically ordered systems. Replacing magnetic ions with non-magnetic analogues removes sites that can participate in magnetic exchange. This has a destabilizing effect on long range order (LRO) which reduces ordering temperature in a fashion dictated by the dimensionality and length scale of the interactions. When dealing with frustrated systems, dilution often produces a spin glass which is a disordered meta-stable state which forms below a well defined freezing temperature $T_{\text{f}}$. In Ga$^{3+}$ diluted Zinc Chromate ZnCr$_{2-x}$Ga$_{x}$O$_{4}$, the AFM state gives way to a spin glass at $x=0.2$~\cite{Lee2008} ($x=0.4$~\cite{Fiorani1984} for bulk magnetic probes).

This work focuses on the effects of magnetic dilution and applied pressure on the ordering behaviour of GeNi$_{2}$O$_{4}$. Specifically, on the magnetic and physical properties of polycrystalline samples of GeNi$_{2-x}$Mg$_{x}$O$_{4}$ ($x=0$ to 1.0). The phase diagram of this family of compounds is determined using bulk probes. Particularly, we seek to understand how these contributions will effect each transition, as they are linked to particular planes (kagom\'{e} or triangular). Results contain herein indicate GeNi$_{2-x}$Mg$_{x}$O$_{4}$ exists in three distinct magnetic phases: antiferromagnetic ($0\leq x \leq 0.05$), ill-defined (x=0.10 and 0.15), and spin glass (x $\geq$ 0.30). The ordered kagom\'{e} and triangular planes, associated with the AFM state, are extremely vulnerable to magnetic dilution as indicated by large percolation thresholds. We suggest that these anomalous values can be explained by a coupled network effect which causes early failure in the LRO state. This weakness to dilution is contrasted by how robust $T_{\text{N1}}$ and $T_{\text{N2}}$ are to applied pressure. In the spin glass regime, $T^{2}$ dependence of the magnetic specific heat reveals a 3D to 2D shift in the behavior of the system due to dilution. 
\section{Experimental}
Polycrystalline samples of GeNi$_{2-x}$Mg$_{x}$O$_{4}$ (x = 0, 0.01, 0.02, 0.03, 0.04, 0.05, 0.10, 0.15, 0.30, 0.45, 0.60, 0.80, 1.0) were formed using solid state synthesis. High purity powders of NiO, MgO and GeO$_{2}$ were mixed and wet milled in acetone. The mixture was placed in an alumina crucible and fired 1373 K for 12 hours. This was repeated, followed by pellet formation, and a heat treatment of 1473 K for 4 hours. The resulting compound was a light green color which became lighter for diluted samples. 

DC susceptibility measurements were taken using a Quantum Design Magnetic Property Measurement System (MPMS). Specific heat and AC susceptibility data were obtained using a Quantum Design Physical Property Measurement System (PPMS). 

High pressure measurements were taken using a custom built pressure cell which is capable of reaching 1.2 GPa. Degassed glycerine was used as a pressure medium and a small piece high purity lead (99.999\%) as a manometer. Internal cell pressure was determined by measuring the superconducting transition of the lead,\cite{Eiling1981} which was placed in a teflon capsule along with the sample and pressure medium.

\section{Results}

\subsection{X Ray Diffraction}
\label{sec:XRD}


X ray powder diffraction data was refined to determine phase formation and lattice 
constants. No impurity phase or starting material Bragg peaks were observed. Refinement 
data is found in Table \ref{tab:main}. There is an observed expansion of lattice 
parameter with increased dilution that does not appear to be linear in nature. An 
expansion of lattice parameter for the substitution of Mg with Ni is expected as the 
crystal radii of Mg and Ni are 0.71 \AA and 0.63 \AA respectively.\cite{Shannon1976} 
However, the non-linear response has be theoretically predicted in systems with 
$\alpha = 0.88$ where $\alpha$ is the ratio of ionic radii ($\alpha=0.89$ for 
Mg$^{2+}$ and Ni$^{2+}$ in octahedral environments).~\cite{Denton1991} At dilution 
levels of approximately 55\%, the material crystallizes in both olivine and spinel 
forms, with full Mg$^{2+}$ substitution in the spinel structure occurring when high 
pressure synthesis techniques are used.~\cite{Navrotsky1976}

\subsection{DC Susceptibility}
\label{sec:DC}
DC susceptibility data have been reported in literature for GeNi$_{2}$O$_{4}$, but 
inconsistencies exist between authors. For example, ordering temperatures T$_{\text{N1}}$ 
and T$_{\text{N2}}$ of 11.4K and 12.0K are agreed upon; however, various Currie-Weiss (CW) 
temperature $\Theta_{\text{cw}}$ values such as -15 K\cite{Diaz2006}, -8.7 K\cite{Lashley2008}, 
-4.4 K\cite{Crawford2003} are reported. Our measurements of GeNi$_{2}$O$_{4}$ successfully 
reproduced the CW results obtained by Diaz \textit{et al} through the addition of a diamagnetic 
term $\chi_{\text{dia}} = 7.20 \times 10^{-5}$ emu/mol$_{\text{NGO}}$ and a temperature 
independent paramagnetic (TIP) term $\chi_{\text{ind}} = 7.42 \textsc{x} 10^{-4}$ 
emu/mol$_{\text{NGO}}$. This TIP value is reasonable for Ni$^{2+}$ in an octahedral 
crystal field with a splitting of 8000 cm$^{-1}$:~\cite{Carlin1986}
 
\begin{equation}
\chi_{\text{ind}} \propto \dfrac{4}{\Delta} \label{equ:TIP}
\end{equation}

\noindent where $\Delta$ is the wavenumber splitting to the first excited state, 
which is large enough to not be thermally populated. These contributions were scaled
appropriately for all of the diluted samples.

\begin{table}
\caption{Summary of data on samples of parent and diluted NGO. Dilution (\%) is 
Mg concentration, superscripts indicate n (N\'{e}el), f (spin glass), + (ill defined) transitions.
 Results from Rietveld refinement are given as lattice constant $a$ and the corresponding $\chi^{2}$.\label{tab:main}}

\begin{center}

\begin{ruledtabular}
\begin{tabular}{ccccc}
Dilution (\%) & $T_{\text{n,f}}$ (K) & $\theta_{\text{CW}}$ (K) & $a$ (\AA)& $\chi^{2}$\\
\hline

0 & $^{n}$11.41, $^{n}$12.01 & 15.0 $\pm$ 0.5 &  8.218(1) & 1.25\\
2.5 & $^{n}$10.89 & 14.8 $\pm$ 0.2 & 8.219(0) & 1.36\\
5.0 & $^{+}$9.0 & 15.1 $\pm$ 0.2 &  8.219(0) & 1.6\\
7.5 & $^{+}$7.76 & 13.5 $\pm$ 0.3 & 8.220(1) & 1.74\\
15& $^{f}$7.41 & 9.4 $\pm$ 0.3 & 8.221(7) & 1.27\\
22.5 & $^{f}$7.03 & 9.2 $\pm$ 0.3 & 8.221(5) & 1.3\\
30 & $^{f}$6.75 & 9.1 $\pm$ 0.2 & 8.221(7) &1.35\\
40 & $^{f}$6.14 & 7,52 $\pm$ 0.09 & 8.222(8) & 1.22\\
50 & $^{f}$5.35 & 6.40 $\pm$ 0.07 & 8.227(4)  & 1.20\\
\end{tabular}
\end{ruledtabular}
\end{center}
\end{table}

Results from DC magnetic data analysis are given in Table \ref{tab:main}. One can identify
two regions ($0.15 \leq x \leq 0.6$ and $0.60 \leq x \leq 1.0$) of linear behavior in 
$\theta_{\text{CW}}$ with respect to dilution. Extrapolating a linear fit line to full dilution 
yields $\theta_{\text{CW}}$=0 within associated error. A similar kink in $\theta_{\text{CW}}$(p) has 
been observed in other diluted frustrated systems, such as 
SrGa$_{12-x}$Cr$_{x}$O$_{19}$ (SGCO(x)).\cite{Martinez1992} 

In spin glasses, when the system is cooled below the freezing temperature, it freezes 
in a metastable state. This instability can be probed using thermoremenant magnetization 
(M$_{TRM}$), which measures the decay of the magnetic susceptibility with time. Our 
cooling and field routine consists of applying a 2000 Oe field at 300 K. The system 
is cooled to 2 K over 12000 seconds. The field is then turned off following a wait time 
of 600 seconds. Once the applied field is removed, the magnetization is measured for 10000 seconds. 

\begin{figure}[hf]
 \includegraphics[width=0.45\textwidth]{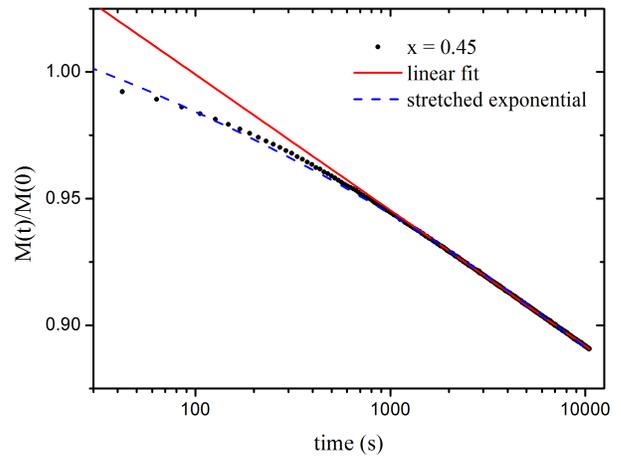}
 \caption{[color online] M$_{\text{TRM}}$ data for GeNi$_{1.55}$Mg$_{0.45}$O$_{4}$ 
 taken over 10000 seconds after a wait time of 600 seconds. The dashed line 
 is the resulting fit using a stretched exponential. The solid line is a linear 
 fit that is constrained to the linear portion of the data.\label{fig:TRM}}
 \end{figure}
 
\begin{equation}
M_{\text{TRM}}=M_{0}e^{-(\frac{t}{\tau})^{1-n}}
\label{equ:strexp}
\end{equation}

The decay of the resulting moment for GeNi$_{1.55}$Mg$_{0.45}$O$_{4}$ is illustrated in Fig. \ref{fig:TRM} and is normalized 
to the initial value measured at t = 0. A stretched exponential function with stretching 
coefficient 1-n was used to fit the M$_{TRM}$. In structural glass theory it is known as 
the Kohlrausch-Williams-Watts (KWW) law and it is used to describe relaxation in physical 
properties of super-cooled liquids.~\cite{Williams1970,Klinger2013} In the 
context of magnetic glasses, this expression has been applied above and below 
$T_{\text{g}}$~\cite{Continentino1986} with a temperature dependence of n near 
$T_{\text{g}}$.~\cite{Chamberlin1984} For NGO spin glasses, the values obtained for n are 
equal within associated error and have a mean of n=0.861 with a standard deviation 
of 0.002. This value falls outside the boundaries predicted using mean field percolation 
in glassy systems.~\cite{Lois2009} Departure from this mean field prediction is not 
unusual, results from structural glass formers show that small values of the 
stretching parameter 1-n correspond to fragile glasses. Physically, n is a measure 
of broadness in the relaxation spectrum, which results in a deviation from Arrhenius 
behavior.~\cite{Ngai1984}

\subsection{Heat Capacity}
The specific heat was measured for diluted samples (x=0 to 0.30) in the temperature range of 
2 K to 300 K. The total heat capacity can be decomposed into the sum of lattice, magnetic and 
electronic contributions. In NGO, the electronic contribution is negligible and can be ignored. 
To obtain the lattice contribution a common approach is to use the low temperature limit 
($T < \theta_{\text{D}}$/50) of the Debye model.~\cite{Gopal1966} In NGO, particularly in the 
diluted samples, the ordering phenomena requires the addition of magnetic correlations just 
above $T_{\text{N}}$ and both gapped and ungapped spin wave contributions to the heat 
capacity.~\cite{Lashley2008} One would require an accurate description of the dilution response 
to these spin wave contributions to yield a meaningful lattice heat capacity. In light of this, 
an alternative method based on Pad\'{e} approximants,~\cite{Goetsch2012} which was employed 
here, allows one to fit convenient regions of the specific heat to obtain the lattice 
contribution. The total heat capacity can be expressed as:

\begin{equation}
C_{\text{p}} =  C_{\text{mag}} + A*C_{\text{pad\'{e}}} + B*C_{\text{E1}} + C*C_{\text{E2}}\label{equ:HCfull}
\end{equation}

where the last three terms model the lattice contribution and the sum of $A$,$B$ and $C$ is 7. 
The Pad\'{e} approximant method is designed to reproduce the Debye function through all temperature
ranges. It therefore utilizes the same Debye assumptions\cite{Gopal1966} (harmonic approximation
and isotropic spin waves) and represents only the acoustic modes. To account for the optical modes,
two Einstein terms $C_{\text{E1}}$ and $C_{\text{E2}}$, have been added, which take the form of 
equation \ref{equ:Eins}. $T_{\text{E}}$ is the corresponding Einstein temperature and R is the gas constant.

\begin{equation}
C_{E}=3R\left(\frac{T_{\text{E}}}{T}\right)^{2}\dfrac{e^{T_{\text{E}}/T}}{(e^{T_{\text{E}}/T}-1)^2}\label{equ:Eins}
\end{equation}

\begin{figure}[hf]
 \includegraphics[width=0.45\textwidth]{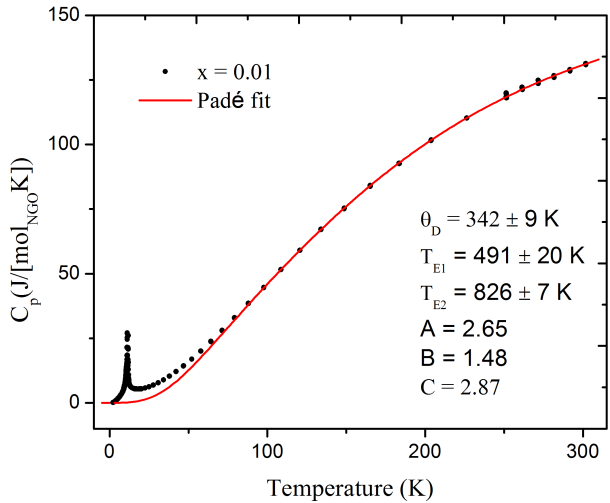}
 \caption{[color online] Specific heat of Ni$_{1.99}$Mg$_{0.01}$GeO$_{4}$. The data is fitted 
 from 120 K to 300 K to the last three terms of equation \ref{equ:HCfull} with the fit line 
 extrapolated to 0 K.\label{fig:padefit}}
 \end{figure}

Figure \ref{fig:padefit} illustrates experimental data for Ni$_{1.99}$Mg$_{0.01}$GeO$_{4}$ and 
the resulting fit using the last three terms of equation \ref{equ:HCfull}. The temperature range 
for the fit is 100 K to 300 K, which was selected to be sufficiently far away from the magnetic 
transitions such that they have a negligible contribution to the specific heat. 

The Debye temperature obtained is $\Theta_{D}$ 342 $\pm$ 9K, which is smaller than the value of 386 K 
found by Lashley \textit{et al}.\cite{Lashley2008} Possible reasons for this difference could be their 
usage of the low temperature Debye form to temperatures as high as 75 K. Due to the approximations made 
in the Debye model, the value of $\Theta_{\text{D}}$ varies with temperature. This variation is 
minimized~\cite{Gopal1966} in the temperature regions $T < \theta_{\text{D}}$/50 (7.72 K) and 
$T > \theta_{\text{D}}$/2 (193 K) if the 386 K value is assumed.

 \begin{figure}[hf]
 \includegraphics[width=0.45\textwidth]{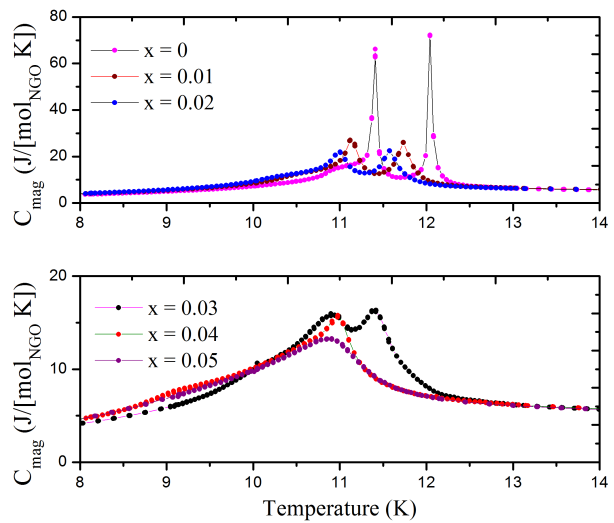}
 \caption{[color online] Plot of the magnetic contribution to specific heat for several diluted samples 
 of GeNi$_{2-x}$Mg$_{x}$O$_{4}$.\label{fig:splithc}}
 \end{figure}

With the lattice contribution determined, the magnetic contribution can be calculated and is 
illustrated in Fig. \ref{fig:splithc}. There is a reduction and broadening of the peaks in the 
specific heat, with increasing dilution of the the Ni$^{2+}$ ions, which is evidence for a shift 
from long to short range ordering. We observe the disappearance of $T_{\text{N2}}$ at 2.0$\%$ 
($x=0.04$) dilution while $T_{\text{N1}}$ persists in a visible but broadened form. This indicates 
the destruction of long range order upon random site dilution occurs first in the triangular planes 
followed by the kagom\'{e} planes. This is interesting since one would expect dilution to affect 
kagom\'{e} ordering to a greater degree as there are three times the number of kagom\'{e} sites 
compared to triangular sites. A Mg$^{2+}$ ion is three times more likely to sit on these planes, 
if the dilution is completely random amongst B sites. This result suggests a heightened importance 
of the linkage role that the kagom\'{e} sites, which are sandwiched between to triangular planes, 
provide to the network of triangular planes.

\subsection{AC Susceptibility}

\begin{figure}[hf]
 \includegraphics[width=0.45\textwidth]{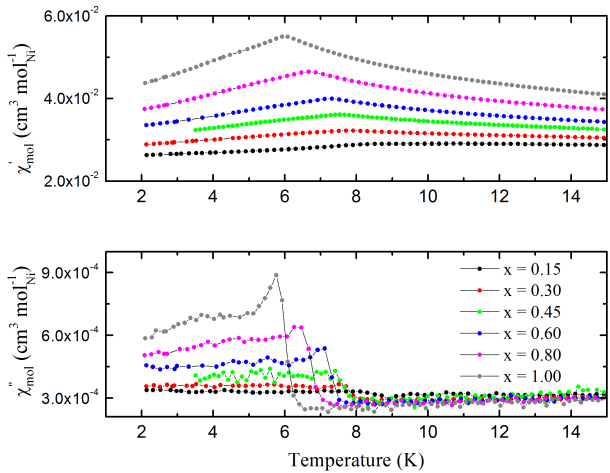}
 \caption{[color online] Plot of the AC susceptibility of diluted samples taken at 7000 Hz.The 
 cusp in the real part and jump in the imaginary part of the dynamic susceptibility indicate 
 spin glass transitions.\label{fig:7000ACMS}}
 \end{figure}

With the destruction of long range order, the dynamic susceptibility becomes an important probe for 
determining the nature of the low temperature state. AC susceptibility measurements were completed 
on samples in the dilution range of $x=0.15$ to 1.00 at 500, 1000, 5000 and 7000 Hz. Data for these 
dilution values, taken at 7000 Hz, is shown in Fig. \ref{fig:7000ACMS}. The onset of canonical spin 
glass behavior can be seen in dilution levels at, and above, 15\%. In canonical spin glasses the 
freezing temperature $T_{\text{f}}$ can be defined as the cusp in the real part of the AC 
susceptibility $\chi^{'}$ and a spike in the imaginary part of the susceptibility $\chi^{''}$. 
This behavior is due to the inability of spins to keep up with the oscillating applied field, 
which diminishes the in phase component $\chi^{'}$, and causes a non-zero out of phase component $\chi^{''}$. 

In diluted NGO, there is a decrease in freezing temperature $T_{\text{f}}$ with increased Mg$^{2+}$ 
concentration, which follows a nearly linear response in the spin glass regime ($0.15<x<1.00$). 
Linear extrapolation of the data yields a non physical result of approximately $x=3$ for the 
intercept on the temperature axis.

An observed frequency dependence on $T_{\text{f}}$ exists and is a characteristic feature of spin 
glasses and systems with broad distributions of energy levels.~\cite{Malinowski2011,Binder1986} 
One way to compare this frequency dependence between different systems is to calculate $\delta$, 
which is a measure of the drift of $T_{\text{f}}$ with frequency. This is often defined as:

\begin{equation}
\delta = \frac{\Delta T_{\text{f}}}{T_{\text{f}}\Delta ln(w)}
\label{equ:sigma}
\end{equation}

\noindent where $w=2\pi f$ and $f$ is the AC driving frequency. The value of sigma can be used to 
determine the relative strength of interaction between the spin entities (clusters or single spins) of 
the spin glass.\cite{Malinowski2011,Dormann1988} For weakly interacting systems, such as 
La$_{0.994}$Gd$_{0.006}$Al$_{2}$,~\cite{Tholence1984} $\delta$ takes a value of 0.056. Conversely, 
for systems with stronger interactions such as canonical spin glasses, $\delta$ has an approximate 
value of 0.002~\cite{Dormann1988}. In Ni$_{2-x}$Mg$_{x}$GeO$_{4}$, $\delta$ takes values: 0.0061, 
0.0074, 0.0065, 0.0084, 0.0091 for dilution levels of $x$ = 0.30, 0.45, 0.60, 0.80, 1.00 respectively. 

The frequency dependence of the spin glass freezing temperature can also be modelled using Arrhenius type, 
Vogel-Fulcher-Tamman law, power law, and stretched exponential.~\cite{Souletie1985} In the Arrhenius 
model ($\tau$ is the inverse of AC driving frequency):

\begin{equation}
\tau = \tau _{0}e^{\frac{E_{\text{a}}}{k_{\text{b}}T_{\text{f}}}}
\label{equ:Arrh}
\end{equation}

\noindent one often finds that equation \ref{equ:Arrh} yields non-physical results for the characteristic 
relaxation time $\tau_{0}$ and activation energy $E_{\text{a}}$.~\cite{Dormann1988} This deviation from 
Arrhenius behavior is a characteristic of a fragile glass former. A modification of the Arrhenius form 
yields the Vogel-Fulcher-Tamman (VFT) law, which was developed in structural glass literature to describe the 
temperature dependence of viscosity.~\cite{Shtrikman1981} It can be applied in systems with a broad 
distributions of relaxation times.

\begin{equation}
\tau = \tau _{0}e^{\frac{E_{\text{a}}}{k_{\text{b}}(T_{\text{f}}-T_{\text{0}})}}
\label{equ:VFT}
\end{equation}

\begin{figure}[hf]
 \includegraphics[width=0.45\textwidth]{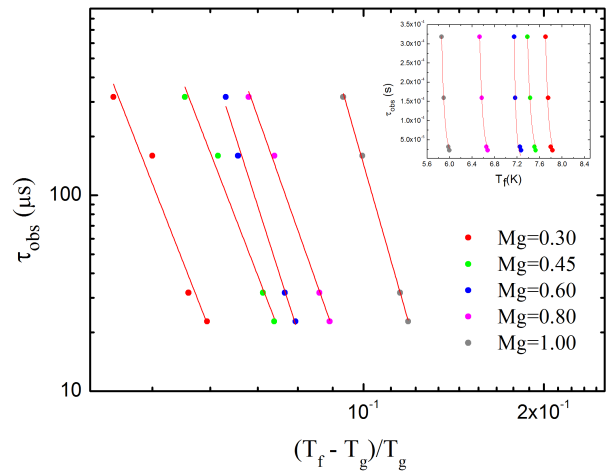}
 \caption{[color online] Plot of AC freezing temperature data fitted to equation \ref{equ:power}. 
 Inset graph illustrates the same data fitted using the VFT equation. \label{fig:power}}
 \end{figure}

Here $T_{0}$ is a measure of interaction strength between spin entities (clusters or single spins) and $E_{a}$ is 
the activation energy. This form, although useful, creates ambiguity as $T_{0}$ is assigned in an 
ad-hoc manner to obtain what is believed to be physically relevant values for $ \tau_{0}$.~\cite{Souletie1985} 
To remedy this issue, one can utilize a power law form reminiscent of expressions used in other 
critical phenomena:

\begin{equation}
\tau = \tau_{*} \left(\frac{T_{f}-T_{g}}{T_{g}}\right) ^{-zv}
\label{equ:power}
\end{equation}

where $T_{\text{f}}$, $\tau_{*}$, $T_{\text{f}}$ are the frequency dependent freezing temperature, 
the characteristic relaxation time and the freezing temperature as $f \rightarrow 0$. Rather than 
fitting all three parameters simultaneously, $T_{\text{f}}$ was determined by the peak in the DC 
susceptibility. Observed data and the resulting fits to equations \ref{equ:VFT} and \ref{equ:power} are 
shown in Fig. \ref{fig:power} with values for the fitted parameters found in Table \ref{tab:AC}.

\begin{table}
\caption{Summary of AC susceptibility fit data on diluted samples.\label{tab:AC}}

\begin{center}

\begin{ruledtabular}
\begin{tabular}{c|c|c|c|c}
Dilution (\%) & zv  & $\tau_{*} (10^{-15} s)$ & $E_{a} (eV)$ & $T_{0} (K)$\\
\hline
15   & 7.8 $\pm$ 1.0  & 3.2 $\pm$ 0.3  & 2.8 $\pm$ 0.2  & 6.44 $\pm$ 0.09\\
22.5 & 7.92 $\pm$ 0.6 & 19  $\pm$ 3    & 2.7 $\pm$ 0.2  & 6.03 $\pm$ 0.08\\
30   & 9.6 $\pm$ 0.6  & 0.50$\pm$ 0.02 & 1.88$\pm$ 0.06 & 6.34 $\pm$ 0.03\\
40   & 8.7 $\pm$ 0.3  & 17 $\pm$ 0.9   & 2.6 $\pm$ 0.1  & 5.28 $\pm$ 0.04\\
50   & 10.7 $\pm$ 0.3 & 2.8 $\pm$ 0.07 & 2.8 $\pm$ 0.1  & 4.56 $\pm$ 0.03\\
\end{tabular}
\end{ruledtabular}
\end{center}
\end{table}

The characteristic relaxation time of the spin glass system $\tau_{*}$ ranges from $10^{-14}$ to $10^{-16}$ 
seconds, which indicative of canonical spin glass behaviour. Contrast this with the so-called cluster glass 
systems like La$_{1.85}$Sr$_{0.15}$Cu$_{1-y}$Ni$_{y}$O$_{4}$, Li$_{x}$Ni$_{2-x}$O$_{2}$ and 
CaBaFe$_{4-x}$Li$_{x}$O$_{7}$, which have $\tau_{*}$ values of $10^{-9.5}$, $7.5\times 10^{-11}$ and $4.9\times 10^{-12}$ 
seconds respectively.\cite{Vijay2009,Sow2013,Malinowski2011} The values of the critical exponent pair $zv$ 
obtained vary from 7.8 to 10.7 and fall within the range 5 to 11 found for several spin glass compounds.~\cite{Souletie1985}

\subsection{High Pressure Effects}

\begin{figure}[h]
 \includegraphics[width=0.45\textwidth]{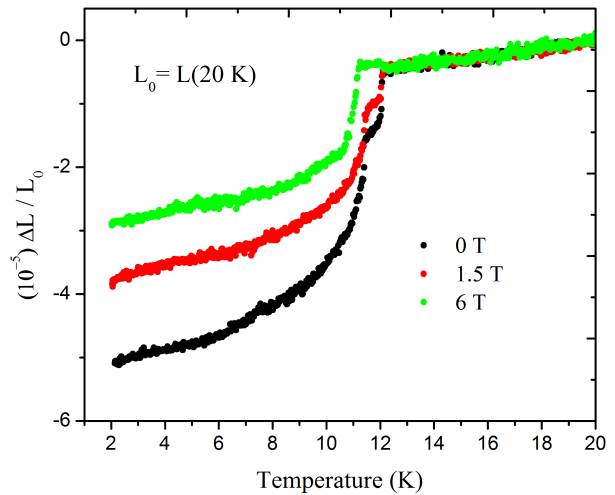}
 \caption{[colour online] Strain $\Delta L\backslash$L$_{0}$, taken at various applied fields, where L$_{0}$ 
 is the reference length at 20 K. The onset of magnetic transitions are denoted by a rapid change in strain.\label{fig:TE1}}
 \end{figure}

Lattice distortion is an important process that can relieve frustration. This is evident particularly 
in the related spinel GeCo$_{2}$O$_{4}$, which undergoes a simultaneous cubic to tetragonal distortion and 
antiferromagnetic transition, where $c$/$a$ $\approx$ 1.0014.~\cite{Hoshi2007} To the authors' knowledge, there 
has been no evidence of a structural transition in NGO using neutron or synchrotron x ray 
probes.~\cite{Crawford2003} The relation between lattice constant and the corresponding magnetic properties 
of NGO can be further tested using high precision thermal expansion measurements. The results are illustrated 
in Fig. \ref{fig:TE1} where the reference length L$_{0}$ is $L(T=20)$. The two AFM transitions are clearly 
indicated by a steep decrease in strain $\Delta L\backslash L_{0}$ indicating lattice constant shrinkage. 
This decrease is affected by applied magnetic field and is most noticeable at the kagom\'{e} ordering 
temperature T$_{\text{N1}}$. Specifically, the change in strain during ordering at T$_{\text{N1}}$ is 
$-6.899\times 10^{-6}$, $-4.943\times 10^{-6}$, and $1.082\times 10^{-6}$ for fields of 0, 1.5, and 6.0 T respectively. 
A similar field induced transition in the nature of TE response has been observed\cite{Reuvekamp2014} in EuTiO$_{3}$
where a cross over between contraction and expansion occurred at 0.675 T, with destruction of long range order at 1 T. 
Conversely, NGO is very robust against applied fields with a spin flop transitions at $H_{1}=30$ T and 
$H_{2}$=37 T at 4 K.\cite{Diaz2004} 

This result suggests that a spin-lattice coupling occurs in NGO. To further test the effects of perturbations on 
Ni$_{2}$GeO$_{4}$, quasi-hydrostatic high pressure susceptibility measurements were taken on a polycrystalline 
sample in the range 0 to 1.2 GPa. To estimate the effect of applied pressure on the lattice it is noted~\citep{Hofmeister1991} 
that several related spinel compounds have bulk moduli $K$ that lie in between 170 to 210 GPa. Using a value of 200 
GPa the strain can be estimated using $K=-V \Delta P \backslash \Delta V$ where 
$\Delta V \approx 3L^{2} \Delta L$ the maximum strain is $10^{-3}$.

\begin{figure}[h]
 \includegraphics[width=0.45\textwidth]{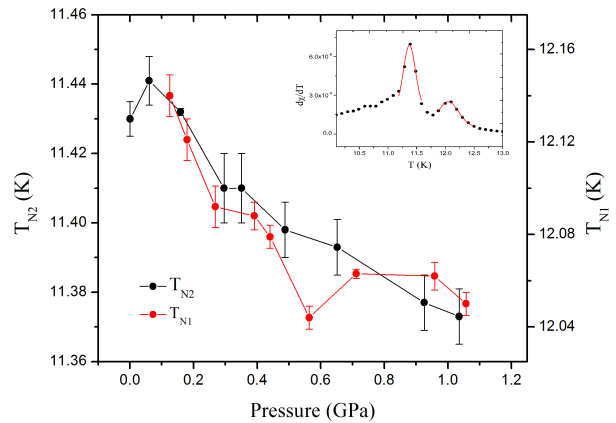}
 \caption{[colour online] Antiferromagnetic transitions $T_{\text{N1}}$ (kagom\'{e}) and $T_{\text{N2}}$ (triangular) 
 in GeNi$_{2}$O$_{4}$ at various applied pressures. The inset graph depicts fits to peaks in 
 d$\chi \backslash$d$T$. \label{fig:pressure}}
 \end{figure}

Figure \ref{fig:pressure} illustrates the effect of applied quasi-hydrostatic pressure on the two AFM 
transitions of polycrystalline NGO. These were determined by fitting Gaussian peak functions to maxima 
in d$\chi \backslash$d$T$, which is shown in the inset. There is a small decrease of similar magnitude in 
the ordering temperatures $T_{\text{N1}}$ and $T_{\text{N2}}$. The observation that both transitions 
have nearly identical responses is somewhat striking for several reasons. First, the kagom\'{e} and 
triangular planes order independently and, as seen in the strain data, yield vastly different lattice effects under applied fields. 
Furthermore, the exchange connectivity for each plane type is different with the kagom\'{e} planes having more nearest-neighbors that reside in the same plane. These results depict a magnetically robust system that can withstand lattice changes 
of varying degree, but still maintain the necessary balance in exchange interactions that yield two AFM transitions.

\section{Discussion}

\begin{figure}[hf]
 \includegraphics[width=0.45\textwidth]{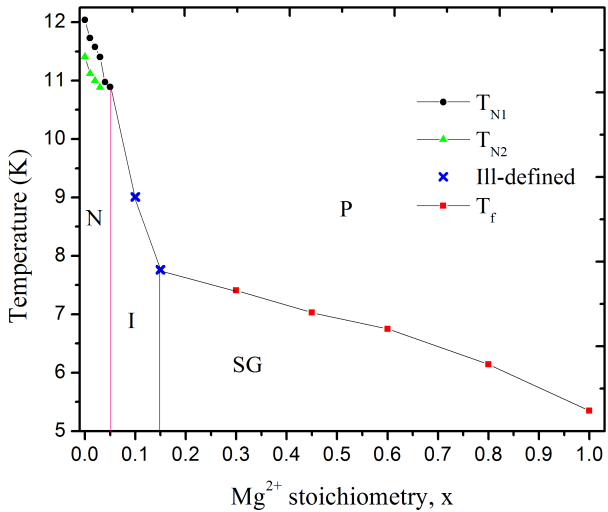}
 \caption{[colour online] Experimental phase diagram of magnetically diluted 
 GeNi$_{2x}$Mg$_{x}$O$_{4}$. There are 4 distinct regions: N - N\'{e}el ordered, I - ill defined, 
 SG - spin glass, and P - paramagnetic.\label{fig:phase4}}
 \end{figure}

Combining the results from specific heat, AC and DC susceptibility measurements, one can 
construct a phase diagram which is shown in Fig. \ref{fig:phase4}. There are 3 distinct 
magnetic phases separated by an indeterminate region, which has no distinct features of a 
spin glass transition or long range ordering. 

\subsection{Low Dilution $x<0.10$}
The effects of B site magnetic dilution appear have a dramatic effect on the two 
long range ordered, low temperature states. The magnitude of this effect varies 
for the different planar types that make up the B site sub-lattice. Figure \ref{fig:phase4} 
illustrates that long range order persists in the kagom\'{e} and triangular planes 
up to dilution values up to x=0.05 and x=0.03 respectively. In this region the N\'{e}el 
temperatures associated with T$_{\text{N1}}$ and T$_{\text{N2}}$ evolve linearly and can be extracted 
to yield the critical dilution level of $x_{\text{d1}}=0.53 \pm 0.07$ and $x_{\text{d2}}=0.70 \pm 0.1$. 
This corresponds to a critical concentration of $p_{\text{c1}}=0.74 \pm 0.04$ for the kagom\'{e} planes 
and $p_{\text{c2}}=0.65 \pm 0.05$ for the triangular planes. The expected 2 dimensional values for 
these planar types with nearest-neighbor interactions are 0.652 (kagom\'{e}) and 0.5 (triangular). 
This is an unusual result as one would expect the experimental values of $p_{\text{c}}$ to be smaller 
than the 2D values. NGO is the antithesis of a 2D system with large interplanar coupling, which is 
believed to give rise to NGO's unique ground state.~\cite{Matsuda2008} Increases in dimensionality 
should reduce p$_{\text{c}}$ as the lattice becomes more connected. For example, the percolation threshold 
for stacked kagom\'{e} and triangular lattices have values~\cite{Marck1997} of 0.3345 and 0.2623
respectively, much lower than 2D analogues. Similarly, if interactions extend beyond nearest 
neighbours, the ordered state becomes very robust with respect to magnetic site removal.~\cite{Fiorani1979}

However, seemingly artificial levels for critical concentration are not unheard of. In the 
kagom\'{e} staircase (Co$_{1-x}$Mg$_{x}$)$_{3}$V$_{2}$O$_{8}$, it was found to have an 
$x_{\text{c}}=0.74$ instead of 0.65 for Ne\'{e}l ordering. This was determined to stem from 
the buckled nature of the kagom\'{e} staircase structure, that quenches exchange interactions 
from the spine sites, which increases $p_{\text{c}}$ from the expected 0.65 to 0.74.~\cite{Fritsch2012} 
In NGO, there is no evidence of major structural distortion,~\cite{Crawford2003} such as buckling, 
that would result in the loss of interactions. Furthermore, NGO is a highly 3D system with large 
spin connectivity due to its large exchange network.

One possibility for this high sensitivity to dilution is the effect of coupling between the spins 
that reside on different plane types. That is, due to the distinct ordering of the kagom\'{e} 
and triangular planes below $T_{\text{N2}}$, one can think of the B site lattice as two, 3D 
interconnected networks: stacked triangular and stacked kagom\'{e}. These networks have connections 
to themselves and to each other. Factoring up to J$_{4}$ with 42 total neighbours, a single spin in 
the triangular network contains 12 links to other triangular sites and 30 links to the kagom\'{e} network. 
Conversely, the kagom\'{e} system is much more self connected with only 10 links to the triangular network 
and 30 to other kagom\'{e} sites. Recent research~\cite{Havlin2014} has shown that coupled networks, 
a network of networks (NON), are more vulnerable to what is often called 'attack', which in this context 
is site dilution. This vulnerability is manifested as an increased percolation threshold~\cite{Buldyrev2010} 
and has been observed in several different network types: Erdos-Renyi, regular random, and 
lattice.~\cite{Bashan2013} Furthermore, with increased internetwork coupling, the more vulnerable the NON 
becomes, which results in larger $p_{\text{c}}$ thresholds. This is in stark contrast to single networks 
where increased connectivity tends to decrease $p_{\text{c}}$.~\cite{Marck1997,Stauffer1994,Buldyrev2010} 
Physically, this can be understood as a diluted site, on either kagom\'{e} or triangular layers, 
affecting both its own layer type as well as the other complementary layer. This perturbation 
could cascade locally through both networks amplifying its effect.

\subsection{Spin Glass ($0.30 \leq x \leq 1.0$)}

A broad specific heat anomaly and a jump in the imaginary component of the AC susceptibility results indicate 
a single spin glass transition that begins at $x=0.30$ (15$\%$ dilution). Somewhat surprisingly, glassiness does 
not form on each plane type, which would presumably give rise to two distinct freezing transitions: 
one for kagom\'{e} and one for triangular spins. Instead there is a single spin glass transition. 
It is unclear if the glass state emerges across the entire B site sub-lattice, or only on the kagom\'{e} 
sites while the triangular planes behave in paramagnetic fashion. Specific heat results suggest the 
latter case due to the apparent destruction of the $T_{\text{N2}}$ in Fig. \ref{fig:splithc}, 
rather than the broadening and merger of both peaks. If the glass state is only composed of kagom\'{e} sites, 
one can compare the above results with structurally similar systems that undergo spin glass transitions.

\begin{figure}[hf]
 \includegraphics[width=0.45\textwidth]{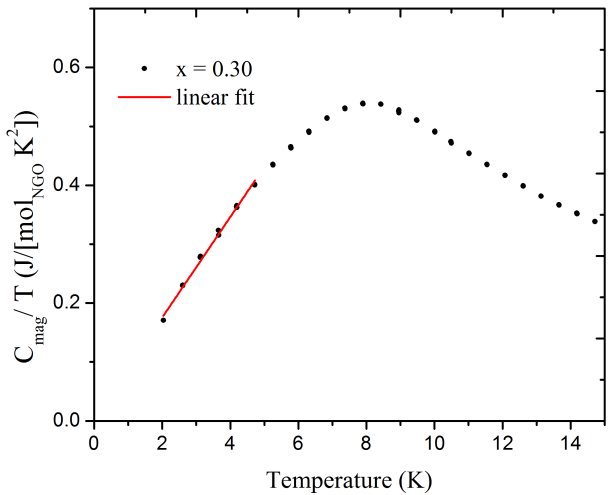}
 \caption{[colour online] Magnetic component of $C_{\text{p}}/T$ for the spin glass 
 GeNi$_{1.7}$Mg$_{0.3}$O$_{4}$. The red linear fit line indicates a $T^{2}$ dependence 
 of the magnetic specific heat at low temperature.\label{fig:LowCp}}
 \end{figure}

One such system, deuteronium jarosite (D$_{3}$O)Fe$_{3}$(SO$_{4}$)$_{2}$(OD)$_{6}$, consists of  Fe$^{3+}$ 
ions (S=5/2) in stacked kagom\'{e} planes, which undergoes a spin glass transition at 13.8 K.~\cite{Wills1998} 
Specific heat results indicate an anomalous low temperature $T^{2}$ behaviour, instead of the linear relationship 
expected for canonical spin glasses. Similar behaviour has been found in SCGO(x) spin glasses.~\cite{Ramirez1990} 
In both cases, it is believed that the $T^{2}$ dependence stems from the strong 2D character of the kagom\'{e} 
planes,~\cite{Ramirez1992} which are the magnetic building blocks of the compounds. In the spin glass 
GeNi$_{1.7}$Mg$_{0.3}$O$_{4}$ there is a low temperature T$^{2}$ dependence, illustrated in Fig. \ref{fig:LowCp}, which could 
indicate a similar 2D character. Furthermore, the $\delta$ value for diluted NGO range from 0.006 to 0.009  
compare favourably to deuteronium jarosite where $\delta=0.01$. In addition, these numbers are comparable the values 
obtained for canonical spin glasses such as CuMn ($\delta=0.002$),~\cite{Mulder1982} and AuMn 
($\delta$=0.0020).~\cite{Mulder1981}

\section{Conclusions}

In this paper the effects of magnetic dilution and spin-lattice coupling on spinel 
type GeNi$_{2}$O$_{4}$ are reported. This was undertaken to probe the nature of the 
long range ordered state of NGO that is formed by two closely spaced magnetic transitions, 
the root of which, is a complex exchange network of 32 relevant neighbours. Thermal expansion 
measurements reveal a field dependent lattice contraction at both $T_{\text{N1}}$ and $T_{\text{N2}}$. 
At an applied magnetic field of 6 T, there is a positive change in strain indicating a field 
induced cross over from lattice contraction to expansion. The phase evolution of this compound 
when Ni$^{2+}$ sites are replaced with non-magnetic Mg$^{2+}$ can be broadly classified into 
three regions: ordered, ill defined, and spin glass.

At dilution levels of $x = 0.30$ and higher the system becomes a spin glass. AC susceptibility and 
thermoremanent magnetization measurements suggest that the system is a canonical type glass. In addition, 
there is no evidence of distinct spin glass transitions for each of the AFM transitions observed in the 
parent compound. It appears that the triangular planes first become disordered, followed by a glass transition 
in the kagom\'{e} planes. The magnetic specific heat in the glassy state has a $T^{2}$ character at low temperature, 
which is unusual for spin glass compounds. This suggests a shift from 3D long range ordered state to a 2D spin glass.

Lastly, the long range ordered state of the kagom\'{e} and triangular planes are extremely susceptible to magnetic 
dilution. Specific heat and magnetic susceptibility measurements indicate that clear signs of magnetic 
ordering disappear for kagom\'{e} spins at approximately 2.5\% Mg$^{2+}$ substitution. Likewise, the 
ordered state for triangular spins appear even more delicate, with a lack of distinct ordering at 1.5$\%$ dilution. 
Through linear fits the percolation thresholds for the kagom\'{e} and triangular planar networks 
were found to be $p_{\text{c1}}=0.74 \pm 0.04$ and $p_{\text{c2}}=0.65 \pm 0.05$ respectively. 
These values are much larger than percolation values for stacked kagom\'{e} ($p_{\text{c}}= 0.3346$) and 
triangular planes ($p_{\text{c}}= 0.2623$). This perplexing result can possibly be explained by treating the 
triangular and kagom\'{e} spins as a network of networks. NONs have been shown to be highly unstable to site 
removal which is amplified by increased connectivity.~\cite{Buldyrev2010} Neutron diffraction experiments 
currently undertaken will hopefully shed light on the nature of the ill-defined phase and the anomalously 
large percolation threshold. If the NON interpretation is correct, to the author's knowledge, NGO could be 
the first reported manifestation of this effect in the solid state. Further research in other magnetic systems 
with distinctly ordered sub-lattices, such as Gd$_{2}$Ti$_{2}$O$_{7}$,\cite{Stewart2004} could be studied 
to find other candidates.

\bibliography{NGO}

\begin{thebibliography}{60}%
\makeatletter
\providecommand \@ifxundefined [1]{%
 \@ifx{#1\undefined}
}%
\providecommand \@ifnum [1]{%
 \ifnum #1\expandafter \@firstoftwo
 \else \expandafter \@secondoftwo
 \fi
}%
\providecommand \@ifx [1]{%
 \ifx #1\expandafter \@firstoftwo
 \else \expandafter \@secondoftwo
 \fi
}%
\providecommand \natexlab [1]{#1}%
\providecommand \enquote  [1]{``#1''}%
\providecommand \bibnamefont  [1]{#1}%
\providecommand \bibfnamefont [1]{#1}%
\providecommand \citenamefont [1]{#1}%
\providecommand \href@noop [0]{\@secondoftwo}%
\providecommand \href [0]{\begingroup \@sanitize@url \@href}%
\providecommand \@href[1]{\@@startlink{#1}\@@href}%
\providecommand \@@href[1]{\endgroup#1\@@endlink}%
\providecommand \@sanitize@url [0]{\catcode `\\12\catcode `\$12\catcode
  `\&12\catcode `\#12\catcode `\^12\catcode `\_12\catcode `\%12\relax}%
\providecommand \@@startlink[1]{}%
\providecommand \@@endlink[0]{}%
\providecommand \url  [0]{\begingroup\@sanitize@url \@url }%
\providecommand \@url [1]{\endgroup\@href {#1}{\urlprefix }}%
\providecommand \urlprefix  [0]{URL }%
\providecommand \Eprint [0]{\href }%
\providecommand \doibase [0]{http://dx.doi.org/}%
\providecommand \selectlanguage [0]{\@gobble}%
\providecommand \bibinfo  [0]{\@secondoftwo}%
\providecommand \bibfield  [0]{\@secondoftwo}%
\providecommand \translation [1]{[#1]}%
\providecommand \BibitemOpen [0]{}%
\providecommand \bibitemStop [0]{}%
\providecommand \bibitemNoStop [0]{.\EOS\space}%
\providecommand \EOS [0]{\spacefactor3000\relax}%
\providecommand \BibitemShut  [1]{\csname bibitem#1\endcsname}%
\let\auto@bib@innerbib\@empty
\bibitem [{\citenamefont {Wiebe}\ and\ \citenamefont
  {Hallas}(2015)}]{Wiebe2015}%
  \BibitemOpen
  \bibfield  {author} {\bibinfo {author} {\bibfnamefont {C.~R.}\ \bibnamefont
  {Wiebe}}\ and\ \bibinfo {author} {\bibfnamefont {A.~M.}\ \bibnamefont
  {Hallas}},\ }\href {\doibase http://dx.doi.org/10.1063/1.4916020} {\bibfield
  {journal} {\bibinfo  {journal} {APL Materials}\ }\textbf {\bibinfo {volume}
  {3}},\ \bibinfo {eid} {041519} (\bibinfo {year} {2015})}\BibitemShut
  {NoStop}%
\bibitem [{\citenamefont {Balents}(2010)}]{Balents2010}%
  \BibitemOpen
  \bibfield  {author} {\bibinfo {author} {\bibfnamefont {L.}~\bibnamefont
  {Balents}},\ }\href {http://dx.doi.org/10.1038/nature08917} {\bibfield
  {journal} {\bibinfo  {journal} {Nature}\ }\textbf {\bibinfo {volume} {464}},\
  \bibinfo {pages} {199} (\bibinfo {year} {2010})}\BibitemShut {NoStop}%
\bibitem [{\citenamefont {Ji}\ \emph {et~al.}(2009)\citenamefont {Ji},
  \citenamefont {Lee}, \citenamefont {Broholm}, \citenamefont {Koo},
  \citenamefont {Ratcliff}, \citenamefont {Cheong},\ and\ \citenamefont
  {Zschack}}]{Ji2009}%
  \BibitemOpen
  \bibfield  {author} {\bibinfo {author} {\bibfnamefont {S.}~\bibnamefont
  {Ji}}, \bibinfo {author} {\bibfnamefont {S.-H.}\ \bibnamefont {Lee}},
  \bibinfo {author} {\bibfnamefont {C.}~\bibnamefont {Broholm}}, \bibinfo
  {author} {\bibfnamefont {T.~Y.}\ \bibnamefont {Koo}}, \bibinfo {author}
  {\bibfnamefont {W.}~\bibnamefont {Ratcliff}}, \bibinfo {author}
  {\bibfnamefont {S.-W.}\ \bibnamefont {Cheong}}, \ and\ \bibinfo {author}
  {\bibfnamefont {P.}~\bibnamefont {Zschack}},\ }\href {\doibase
  10.1103/PhysRevLett.103.037201} {\bibfield  {journal} {\bibinfo  {journal}
  {Phys. Rev. Lett.}\ }\textbf {\bibinfo {volume} {103}},\ \bibinfo {pages}
  {037201} (\bibinfo {year} {2009})}\BibitemShut {NoStop}%
\bibitem [{\citenamefont {Castelnovo}\ \emph {et~al.}(2008)\citenamefont
  {Castelnovo}, \citenamefont {Moessner},\ and\ \citenamefont
  {Sondhi}}]{Castelnovo2008}%
  \BibitemOpen
  \bibfield  {author} {\bibinfo {author} {\bibfnamefont {C.}~\bibnamefont
  {Castelnovo}}, \bibinfo {author} {\bibfnamefont {R.}~\bibnamefont
  {Moessner}}, \ and\ \bibinfo {author} {\bibfnamefont {S.~L.}\ \bibnamefont
  {Sondhi}},\ }\href {http://dx.doi.org/10.1038/nature06433} {\bibfield
  {journal} {\bibinfo  {journal} {Nature}\ }\textbf {\bibinfo {volume} {451}},\
  \bibinfo {pages} {42} (\bibinfo {year} {2008})}\BibitemShut {NoStop}%
\bibitem [{\citenamefont {Fischer}\ and\ \citenamefont
  {Hertz}(1993)}]{Fischer1993}%
  \BibitemOpen
  \bibfield  {author} {\bibinfo {author} {\bibfnamefont {K.}~\bibnamefont
  {Fischer}}\ and\ \bibinfo {author} {\bibfnamefont {J.}~\bibnamefont
  {Hertz}},\ }\href {http://books.google.ca/books?id=zXqel8dS-rIC} {\emph
  {\bibinfo {title} {Spin Glasses}}},\ Cambridge Studies in Magnetism\
  (\bibinfo  {publisher} {Cambridge University Press},\ \bibinfo {year}
  {1993})\BibitemShut {NoStop}%
\bibitem [{\citenamefont {Harris}\ \emph {et~al.}(1997)\citenamefont {Harris},
  \citenamefont {Bramwell}, \citenamefont {McMorrow}, \citenamefont {Zeiske},\
  and\ \citenamefont {Godfrey}}]{Harris1997}%
  \BibitemOpen
  \bibfield  {author} {\bibinfo {author} {\bibfnamefont {M.~J.}\ \bibnamefont
  {Harris}}, \bibinfo {author} {\bibfnamefont {S.~T.}\ \bibnamefont
  {Bramwell}}, \bibinfo {author} {\bibfnamefont {D.~F.}\ \bibnamefont
  {McMorrow}}, \bibinfo {author} {\bibfnamefont {T.}~\bibnamefont {Zeiske}}, \
  and\ \bibinfo {author} {\bibfnamefont {K.~W.}\ \bibnamefont {Godfrey}},\
  }\href {\doibase 10.1103/PhysRevLett.79.2554} {\bibfield  {journal} {\bibinfo
   {journal} {Phys. Rev. Lett.}\ }\textbf {\bibinfo {volume} {79}},\ \bibinfo
  {pages} {2554} (\bibinfo {year} {1997})}\BibitemShut {NoStop}%
\bibitem [{\citenamefont {Binder}\ and\ \citenamefont
  {Young}(1986)}]{Binder1986}%
  \BibitemOpen
  \bibfield  {author} {\bibinfo {author} {\bibfnamefont {K.}~\bibnamefont
  {Binder}}\ and\ \bibinfo {author} {\bibfnamefont {A.~P.}\ \bibnamefont
  {Young}},\ }\href {\doibase 10.1103/RevModPhys.58.801} {\bibfield  {journal}
  {\bibinfo  {journal} {Rev. Mod. Phys.}\ }\textbf {\bibinfo {volume} {58}},\
  \bibinfo {pages} {801} (\bibinfo {year} {1986})}\BibitemShut {NoStop}%
\bibitem [{\citenamefont {Diep}(2013)}]{Diep2013}%
  \BibitemOpen
  \bibfield  {author} {\bibinfo {author} {\bibfnamefont {H.}~\bibnamefont
  {Diep}},\ }\href {https://books.google.ca/books?id=KMTojwEACAAJ} {\emph
  {\bibinfo {title} {Frustrated Spin Systems}}}\ (\bibinfo  {publisher} {World
  Scientific},\ \bibinfo {year} {2013})\BibitemShut {NoStop}%
\bibitem [{\citenamefont {Kang}\ \emph {et~al.}(2014)\citenamefont {Kang},
  \citenamefont {Shan}, \citenamefont {Ko\ifmmode~\check{s}\else
  \v{s}\fi{}mrlj}, \citenamefont {Noorduin}, \citenamefont {Shian},
  \citenamefont {Weaver}, \citenamefont {Clarke},\ and\ \citenamefont
  {Bertoldi}}]{Kang2014}%
  \BibitemOpen
  \bibfield  {author} {\bibinfo {author} {\bibfnamefont {S.~H.}\ \bibnamefont
  {Kang}}, \bibinfo {author} {\bibfnamefont {S.}~\bibnamefont {Shan}}, \bibinfo
  {author} {\bibfnamefont {A.}~\bibnamefont {Ko\ifmmode~\check{s}\else
  \v{s}\fi{}mrlj}}, \bibinfo {author} {\bibfnamefont {W.~L.}\ \bibnamefont
  {Noorduin}}, \bibinfo {author} {\bibfnamefont {S.}~\bibnamefont {Shian}},
  \bibinfo {author} {\bibfnamefont {J.~C.}\ \bibnamefont {Weaver}}, \bibinfo
  {author} {\bibfnamefont {D.~R.}\ \bibnamefont {Clarke}}, \ and\ \bibinfo
  {author} {\bibfnamefont {K.}~\bibnamefont {Bertoldi}},\ }\href {\doibase
  10.1103/PhysRevLett.112.098701} {\bibfield  {journal} {\bibinfo  {journal}
  {Phys. Rev. Lett.}\ }\textbf {\bibinfo {volume} {112}},\ \bibinfo {pages}
  {098701} (\bibinfo {year} {2014})}\BibitemShut {NoStop}%
\bibitem [{\citenamefont {Ramirez}(1994)}]{Ramirez1994}%
  \BibitemOpen
  \bibfield  {author} {\bibinfo {author} {\bibfnamefont {A.~P.}\ \bibnamefont
  {Ramirez}},\ }\href {\doibase 10.1146/annurev.ms.24.080194.002321} {\bibfield
   {journal} {\bibinfo  {journal} {Annual Review of Materials Science}\
  }\textbf {\bibinfo {volume} {24}},\ \bibinfo {pages} {453} (\bibinfo {year}
  {1994})}\BibitemShut {NoStop}%
\bibitem [{\citenamefont {Greedan}(2001)}]{Greedan2001}%
  \BibitemOpen
  \bibfield  {author} {\bibinfo {author} {\bibfnamefont {J.~E.}\ \bibnamefont
  {Greedan}},\ }\href {\doibase 10.1039/B003682J} {\bibfield  {journal}
  {\bibinfo  {journal} {J. Mater. Chem.}\ }\textbf {\bibinfo {volume} {11}},\
  \bibinfo {pages} {37} (\bibinfo {year} {2001})}\BibitemShut {NoStop}%
\bibitem [{\citenamefont {Gardner}\ \emph {et~al.}(2010)\citenamefont
  {Gardner}, \citenamefont {Gingras},\ and\ \citenamefont
  {Greedan}}]{Gardner2010}%
  \BibitemOpen
  \bibfield  {author} {\bibinfo {author} {\bibfnamefont {J.~S.}\ \bibnamefont
  {Gardner}}, \bibinfo {author} {\bibfnamefont {M.~J.~P.}\ \bibnamefont
  {Gingras}}, \ and\ \bibinfo {author} {\bibfnamefont {J.~E.}\ \bibnamefont
  {Greedan}},\ }\href {\doibase 10.1103/RevModPhys.82.53} {\bibfield  {journal}
  {\bibinfo  {journal} {Rev. Mod. Phys.}\ }\textbf {\bibinfo {volume} {82}},\
  \bibinfo {pages} {53} (\bibinfo {year} {2010})}\BibitemShut {NoStop}%
\bibitem [{\citenamefont {Navrotsky}\ and\ \citenamefont
  {Hughes}(1976)}]{Navrotsky1976}%
  \BibitemOpen
  \bibfield  {author} {\bibinfo {author} {\bibfnamefont {A.}~\bibnamefont
  {Navrotsky}}\ and\ \bibinfo {author} {\bibfnamefont {L.}~\bibnamefont
  {Hughes}},\ }\href@noop {} {\bibfield  {journal} {\bibinfo  {journal}
  {Journal of Solid State Chemistry}\ }\textbf {\bibinfo {volume} {16}},\
  \bibinfo {pages} {185} (\bibinfo {year} {1976})}\BibitemShut {NoStop}%
\bibitem [{\citenamefont {Lacroix}\ \emph {et~al.}(2011)\citenamefont
  {Lacroix}, \citenamefont {Mendels},\ and\ \citenamefont
  {Mila}}]{Lacroix2011}%
  \BibitemOpen
  \bibfield  {author} {\bibinfo {author} {\bibfnamefont {C.}~\bibnamefont
  {Lacroix}}, \bibinfo {author} {\bibfnamefont {P.}~\bibnamefont {Mendels}}, \
  and\ \bibinfo {author} {\bibfnamefont {F.}~\bibnamefont {Mila}},\ }\href
  {http://books.google.ca/books?id=utSV09ZuhOkC} {\emph {\bibinfo {title}
  {Introduction to Frustrated Magnetism: Materials, Experiments, Theory}}},\
  Springer Series in Solid-State Sciences\ (\bibinfo  {publisher} {Springer},\
  \bibinfo {year} {2011})\BibitemShut {NoStop}%
\bibitem [{\citenamefont {Lee}\ \emph {et~al.}(2002)\citenamefont {Lee},
  \citenamefont {Broholm}, \citenamefont {Ratcliff}, \citenamefont
  {Gasparovic}, \citenamefont {Huang}, \citenamefont {Kim},\ and\ \citenamefont
  {Cheong}}]{Lee2002}%
  \BibitemOpen
  \bibfield  {author} {\bibinfo {author} {\bibfnamefont {S.-H.}\ \bibnamefont
  {Lee}}, \bibinfo {author} {\bibfnamefont {C.}~\bibnamefont {Broholm}},
  \bibinfo {author} {\bibfnamefont {W.}~\bibnamefont {Ratcliff}}, \bibinfo
  {author} {\bibfnamefont {G.}~\bibnamefont {Gasparovic}}, \bibinfo {author}
  {\bibfnamefont {Q.}~\bibnamefont {Huang}}, \bibinfo {author} {\bibfnamefont
  {T.~H.}\ \bibnamefont {Kim}}, \ and\ \bibinfo {author} {\bibfnamefont
  {S.-W.}\ \bibnamefont {Cheong}},\ }\href
  {http://dx.doi.org/10.1038/nature00964} {\bibfield  {journal} {\bibinfo
  {journal} {Nature}\ }\textbf {\bibinfo {volume} {418}},\ \bibinfo {pages}
  {856} (\bibinfo {year} {2002})}\BibitemShut {NoStop}%
\bibitem [{\citenamefont {Lashley}\ \emph {et~al.}(2008)\citenamefont
  {Lashley}, \citenamefont {Stevens}, \citenamefont {Crawford}, \citenamefont
  {Boerio-Goates}, \citenamefont {Woodfield}, \citenamefont {Qiu},
  \citenamefont {Lynn}, \citenamefont {Goddard},\ and\ \citenamefont
  {Fisher}}]{Lashley2008}%
  \BibitemOpen
  \bibfield  {author} {\bibinfo {author} {\bibfnamefont {J.}~\bibnamefont
  {Lashley}}, \bibinfo {author} {\bibfnamefont {R.}~\bibnamefont {Stevens}},
  \bibinfo {author} {\bibfnamefont {M.}~\bibnamefont {Crawford}}, \bibinfo
  {author} {\bibfnamefont {J.}~\bibnamefont {Boerio-Goates}}, \bibinfo {author}
  {\bibfnamefont {B.}~\bibnamefont {Woodfield}}, \bibinfo {author}
  {\bibfnamefont {Y.}~\bibnamefont {Qiu}}, \bibinfo {author} {\bibfnamefont
  {J.}~\bibnamefont {Lynn}}, \bibinfo {author} {\bibfnamefont {P.}~\bibnamefont
  {Goddard}}, \ and\ \bibinfo {author} {\bibfnamefont {R.}~\bibnamefont
  {Fisher}},\ }\href {\doibase 10.1103/PhysRevB.78.104406} {\bibfield
  {journal} {\bibinfo  {journal} {Physical Review B}\ }\textbf {\bibinfo
  {volume} {78}},\ \bibinfo {pages} {104406} (\bibinfo {year}
  {2008})}\BibitemShut {NoStop}%
\bibitem [{\citenamefont {Yamada}\ \emph {et~al.}(2000)\citenamefont {Yamada},
  \citenamefont {Hiroi}, \citenamefont {Takano}, \citenamefont {Nohara},\ and\
  \citenamefont {Takagi}}]{Yamada2000}%
  \BibitemOpen
  \bibfield  {author} {\bibinfo {author} {\bibfnamefont {T.}~\bibnamefont
  {Yamada}}, \bibinfo {author} {\bibfnamefont {Z.}~\bibnamefont {Hiroi}},
  \bibinfo {author} {\bibfnamefont {M.}~\bibnamefont {Takano}}, \bibinfo
  {author} {\bibfnamefont {M.}~\bibnamefont {Nohara}}, \ and\ \bibinfo {author}
  {\bibfnamefont {H.}~\bibnamefont {Takagi}},\ }\href {\doibase
  10.1143/JPSJ.69.1477} {\bibfield  {journal} {\bibinfo  {journal} {Journal of
  the Physical Society of Japan}\ }\textbf {\bibinfo {volume} {69}},\ \bibinfo
  {pages} {1477} (\bibinfo {year} {2000})}\BibitemShut {NoStop}%
\bibitem [{\citenamefont {Crawford}\ \emph {et~al.}(2003)\citenamefont
  {Crawford}, \citenamefont {Harlow}, \citenamefont {Lee}, \citenamefont
  {Zhang}, \citenamefont {Hormadaly}, \citenamefont {Flippen}, \citenamefont
  {Huang}, \citenamefont {Lynn}, \citenamefont {Stevens}, \citenamefont
  {Woodfield}, \citenamefont {Boerio-Goates},\ and\ \citenamefont
  {Fisher}}]{Crawford2003}%
  \BibitemOpen
  \bibfield  {author} {\bibinfo {author} {\bibfnamefont {M.}~\bibnamefont
  {Crawford}}, \bibinfo {author} {\bibfnamefont {R.}~\bibnamefont {Harlow}},
  \bibinfo {author} {\bibfnamefont {P.}~\bibnamefont {Lee}}, \bibinfo {author}
  {\bibfnamefont {Y.}~\bibnamefont {Zhang}}, \bibinfo {author} {\bibfnamefont
  {J.}~\bibnamefont {Hormadaly}}, \bibinfo {author} {\bibfnamefont
  {R.}~\bibnamefont {Flippen}}, \bibinfo {author} {\bibfnamefont
  {Q.}~\bibnamefont {Huang}}, \bibinfo {author} {\bibfnamefont
  {J.}~\bibnamefont {Lynn}}, \bibinfo {author} {\bibfnamefont {R.}~\bibnamefont
  {Stevens}}, \bibinfo {author} {\bibfnamefont {B.}~\bibnamefont {Woodfield}},
  \bibinfo {author} {\bibfnamefont {J.}~\bibnamefont {Boerio-Goates}}, \ and\
  \bibinfo {author} {\bibfnamefont {R.}~\bibnamefont {Fisher}},\ }\href
  {\doibase 10.1103/PhysRevB.68.220408} {\bibfield  {journal} {\bibinfo
  {journal} {Physical Review B}\ }\textbf {\bibinfo {volume} {68}},\ \bibinfo
  {pages} {220408} (\bibinfo {year} {2003})}\BibitemShut {NoStop}%
\bibitem [{\citenamefont {Matsuda}\ \emph {et~al.}(2008)\citenamefont
  {Matsuda}, \citenamefont {Chung}, \citenamefont {Park}, \citenamefont {Sato},
  \citenamefont {Matsuno}, \citenamefont {Katori}, \citenamefont {Takagi},
  \citenamefont {Kakurai}, \citenamefont {Kamazawa}, \citenamefont {Tsunoda},
  \citenamefont {Kagomiya}, \citenamefont {Henley},\ and\ \citenamefont
  {Lee}}]{Matsuda2008}%
  \BibitemOpen
  \bibfield  {author} {\bibinfo {author} {\bibfnamefont {M.}~\bibnamefont
  {Matsuda}}, \bibinfo {author} {\bibfnamefont {J.~H.}\ \bibnamefont {Chung}},
  \bibinfo {author} {\bibfnamefont {S.}~\bibnamefont {Park}}, \bibinfo {author}
  {\bibfnamefont {T.~J.}\ \bibnamefont {Sato}}, \bibinfo {author}
  {\bibfnamefont {K.}~\bibnamefont {Matsuno}}, \bibinfo {author} {\bibfnamefont
  {H.~A.}\ \bibnamefont {Katori}}, \bibinfo {author} {\bibfnamefont
  {H.}~\bibnamefont {Takagi}}, \bibinfo {author} {\bibfnamefont
  {K.}~\bibnamefont {Kakurai}}, \bibinfo {author} {\bibfnamefont
  {K.}~\bibnamefont {Kamazawa}}, \bibinfo {author} {\bibfnamefont
  {Y.}~\bibnamefont {Tsunoda}}, \bibinfo {author} {\bibfnamefont
  {I.}~\bibnamefont {Kagomiya}}, \bibinfo {author} {\bibfnamefont {C.~L.}\
  \bibnamefont {Henley}}, \ and\ \bibinfo {author} {\bibfnamefont {S.~H.}\
  \bibnamefont {Lee}},\ }\href {\doibase 10.1209/0295-5075/82/37006} {\bibfield
   {journal} {\bibinfo  {journal} {EPL (Europhysics Letters)}\ }\textbf
  {\bibinfo {volume} {82}},\ \bibinfo {pages} {37006} (\bibinfo {year}
  {2008})}\BibitemShut {NoStop}%
\bibitem [{\citenamefont {Ashcroft}\ and\ \citenamefont
  {Mermin}(1976)}]{Ashcroft1976}%
  \BibitemOpen
  \bibfield  {author} {\bibinfo {author} {\bibfnamefont {N.}~\bibnamefont
  {Ashcroft}}\ and\ \bibinfo {author} {\bibfnamefont {N.}~\bibnamefont
  {Mermin}},\ }\href {http://books.google.ca/books?id=1C9HAQAAIAAJ} {\emph
  {\bibinfo {title} {Solid State Physics}}},\ HRW international editions\
  (\bibinfo  {publisher} {Holt, Rinehart and Winston},\ \bibinfo {year}
  {1976})\BibitemShut {NoStop}%
\bibitem [{\citenamefont {Lee}\ \emph {et~al.}(2008)\citenamefont {Lee},
  \citenamefont {Ratcliff}, \citenamefont {Huang}, \citenamefont {Kim},\ and\
  \citenamefont {Cheong}}]{Lee2008}%
  \BibitemOpen
  \bibfield  {author} {\bibinfo {author} {\bibfnamefont {S.-H.}\ \bibnamefont
  {Lee}}, \bibinfo {author} {\bibfnamefont {W.}~\bibnamefont {Ratcliff}},
  \bibinfo {author} {\bibfnamefont {Q.}~\bibnamefont {Huang}}, \bibinfo
  {author} {\bibfnamefont {T.}~\bibnamefont {Kim}}, \ and\ \bibinfo {author}
  {\bibfnamefont {S.-W.}\ \bibnamefont {Cheong}},\ }\href {\doibase
  10.1103/PhysRevB.77.014405} {\bibfield  {journal} {\bibinfo  {journal}
  {Physical Review B}\ }\textbf {\bibinfo {volume} {77}},\ \bibinfo {pages}
  {014405} (\bibinfo {year} {2008})}\BibitemShut {NoStop}%
\bibitem [{\citenamefont {Fiorani}\ \emph {et~al.}(1984)\citenamefont
  {Fiorani}, \citenamefont {Viticoli}, \citenamefont {Dormann}, \citenamefont
  {Tholence},\ and\ \citenamefont {Murani}}]{Fiorani1984}%
  \BibitemOpen
  \bibfield  {author} {\bibinfo {author} {\bibfnamefont {D.}~\bibnamefont
  {Fiorani}}, \bibinfo {author} {\bibfnamefont {S.}~\bibnamefont {Viticoli}},
  \bibinfo {author} {\bibfnamefont {J.~L.}\ \bibnamefont {Dormann}}, \bibinfo
  {author} {\bibfnamefont {J.~L.}\ \bibnamefont {Tholence}}, \ and\ \bibinfo
  {author} {\bibfnamefont {A.~P.}\ \bibnamefont {Murani}},\ }\href {\doibase
  10.1103/PhysRevB.30.2776} {\bibfield  {journal} {\bibinfo  {journal} {Phys.
  Rev. B}\ }\textbf {\bibinfo {volume} {30}},\ \bibinfo {pages} {2776}
  (\bibinfo {year} {1984})}\BibitemShut {NoStop}%
\bibitem [{\citenamefont {Eiling}\ and\ \citenamefont
  {Schilling}(1981)}]{Eiling1981}%
  \BibitemOpen
  \bibfield  {author} {\bibinfo {author} {\bibfnamefont {A.}~\bibnamefont
  {Eiling}}\ and\ \bibinfo {author} {\bibfnamefont {J.~S.}\ \bibnamefont
  {Schilling}},\ }\href {http://stacks.iop.org/0305-4608/11/i=3/a=010}
  {\bibfield  {journal} {\bibinfo  {journal} {Journal of Physics F: Metal
  Physics}\ }\textbf {\bibinfo {volume} {11}},\ \bibinfo {pages} {623}
  (\bibinfo {year} {1981})}\BibitemShut {NoStop}%
\bibitem [{\citenamefont {Shannon}(1976)}]{Shannon1976}%
  \BibitemOpen
  \bibfield  {author} {\bibinfo {author} {\bibfnamefont {R.~D.}\ \bibnamefont
  {Shannon}},\ }\href {\doibase 10.1107/S0567739476001551} {\bibfield
  {journal} {\bibinfo  {journal} {Acta Crystallographica Section A}\ }\textbf
  {\bibinfo {volume} {32}},\ \bibinfo {pages} {751} (\bibinfo {year}
  {1976})}\BibitemShut {NoStop}%
\bibitem [{\citenamefont {Denton}\ and\ \citenamefont
  {Ashcroft}(1991)}]{Denton1991}%
  \BibitemOpen
  \bibfield  {author} {\bibinfo {author} {\bibfnamefont {A.~R.}\ \bibnamefont
  {Denton}}\ and\ \bibinfo {author} {\bibfnamefont {N.~W.}\ \bibnamefont
  {Ashcroft}},\ }\href {\doibase 10.1103/PhysRevA.43.3161} {\bibfield
  {journal} {\bibinfo  {journal} {Phys. Rev. A}\ }\textbf {\bibinfo {volume}
  {43}},\ \bibinfo {pages} {3161} (\bibinfo {year} {1991})}\BibitemShut
  {NoStop}%
\bibitem [{\citenamefont {Diaz}\ \emph {et~al.}(2006)\citenamefont {Diaz},
  \citenamefont {de~Brion}, \citenamefont {Chouteau}, \citenamefont {Canals},
  \citenamefont {Simonet},\ and\ \citenamefont {Strobel}}]{Diaz2006}%
  \BibitemOpen
  \bibfield  {author} {\bibinfo {author} {\bibfnamefont {S.}~\bibnamefont
  {Diaz}}, \bibinfo {author} {\bibfnamefont {S.}~\bibnamefont {de~Brion}},
  \bibinfo {author} {\bibfnamefont {G.}~\bibnamefont {Chouteau}}, \bibinfo
  {author} {\bibfnamefont {B.}~\bibnamefont {Canals}}, \bibinfo {author}
  {\bibfnamefont {V.}~\bibnamefont {Simonet}}, \ and\ \bibinfo {author}
  {\bibfnamefont {P.}~\bibnamefont {Strobel}},\ }\href {\doibase
  10.1103/PhysRevB.74.092404} {\bibfield  {journal} {\bibinfo  {journal}
  {Physical Review B}\ }\textbf {\bibinfo {volume} {74}},\ \bibinfo {pages}
  {092404} (\bibinfo {year} {2006})}\BibitemShut {NoStop}%
\bibitem [{\citenamefont {Carlin}(1986)}]{Carlin1986}%
  \BibitemOpen
  \bibfield  {author} {\bibinfo {author} {\bibfnamefont {R.~L.}\ \bibnamefont
  {Carlin}},\ }\href@noop {} {\emph {\bibinfo {title} {Magnetochemistry}}}\
  (\bibinfo  {publisher} {Springer-Verlag},\ \bibinfo {year}
  {1986})\BibitemShut {NoStop}%
\bibitem [{\citenamefont {Mart\'{i}nez}\ \emph {et~al.}(1992)\citenamefont
  {Mart\'{i}nez}, \citenamefont {Sandiumenge}, \citenamefont {Rouco},
  \citenamefont {Labarta}, \citenamefont {Rodr\'{i}guez-Carvajal},
  \citenamefont {Tovar}, \citenamefont {Causa}, \citenamefont {Gal\'{i}},\ and\
  \citenamefont {Obradors}}]{Martinez1992}%
  \BibitemOpen
  \bibfield  {author} {\bibinfo {author} {\bibfnamefont {B.}~\bibnamefont
  {Mart\'{i}nez}}, \bibinfo {author} {\bibfnamefont {F.}~\bibnamefont
  {Sandiumenge}}, \bibinfo {author} {\bibfnamefont {A.}~\bibnamefont {Rouco}},
  \bibinfo {author} {\bibfnamefont {A.}~\bibnamefont {Labarta}}, \bibinfo
  {author} {\bibfnamefont {J.}~\bibnamefont {Rodr\'{i}guez-Carvajal}}, \bibinfo
  {author} {\bibfnamefont {M.}~\bibnamefont {Tovar}}, \bibinfo {author}
  {\bibfnamefont {M.}~\bibnamefont {Causa}}, \bibinfo {author} {\bibfnamefont
  {S.}~\bibnamefont {Gal\'{i}}}, \ and\ \bibinfo {author} {\bibfnamefont
  {X.}~\bibnamefont {Obradors}},\ }\href {\doibase 10.1103/PhysRevB.46.10786}
  {\bibfield  {journal} {\bibinfo  {journal} {Phys. Rev. B}\ }\textbf {\bibinfo
  {volume} {46}},\ \bibinfo {pages} {10786} (\bibinfo {year}
  {1992})}\BibitemShut {NoStop}%
\bibitem [{\citenamefont {Williams}\ and\ \citenamefont
  {Watts}(1970)}]{Williams1970}%
  \BibitemOpen
  \bibfield  {author} {\bibinfo {author} {\bibfnamefont {G.}~\bibnamefont
  {Williams}}\ and\ \bibinfo {author} {\bibfnamefont {D.~C.}\ \bibnamefont
  {Watts}},\ }\href {\doibase 10.1039/TF9706600080} {\bibfield  {journal}
  {\bibinfo  {journal} {Trans. Faraday Soc.}\ }\textbf {\bibinfo {volume}
  {66}},\ \bibinfo {pages} {80} (\bibinfo {year} {1970})}\BibitemShut {NoStop}%
\bibitem [{\citenamefont {Klinger}(2013)}]{Klinger2013}%
  \BibitemOpen
  \bibfield  {author} {\bibinfo {author} {\bibfnamefont {M.}~\bibnamefont
  {Klinger}},\ }\href {http://books.google.ca/books?id=TOX\_ugAACAAJ} {\emph
  {\bibinfo {title} {Glassy Disordered Systems: Glass Formation and Universal
  Anomalous Low-energy Properties}}}\ (\bibinfo  {publisher} {World
  Scientific},\ \bibinfo {year} {2013})\BibitemShut {NoStop}%
\bibitem [{\citenamefont {Continentino}\ and\ \citenamefont
  {Malozemoff}(1986)}]{Continentino1986}%
  \BibitemOpen
  \bibfield  {author} {\bibinfo {author} {\bibfnamefont {M.~A.}\ \bibnamefont
  {Continentino}}\ and\ \bibinfo {author} {\bibfnamefont {A.~P.}\ \bibnamefont
  {Malozemoff}},\ }\href {\doibase 10.1103/PhysRevB.33.3591} {\bibfield
  {journal} {\bibinfo  {journal} {Phys. Rev. B}\ }\textbf {\bibinfo {volume}
  {33}},\ \bibinfo {pages} {3591} (\bibinfo {year} {1986})},\ \bibinfo {note}
  {mTRM}\BibitemShut {NoStop}%
\bibitem [{\citenamefont {Chamberlin}\ \emph {et~al.}(1984)\citenamefont
  {Chamberlin}, \citenamefont {Mozurkewich},\ and\ \citenamefont
  {Orbach}}]{Chamberlin1984}%
  \BibitemOpen
  \bibfield  {author} {\bibinfo {author} {\bibfnamefont {R.~V.}\ \bibnamefont
  {Chamberlin}}, \bibinfo {author} {\bibfnamefont {G.}~\bibnamefont
  {Mozurkewich}}, \ and\ \bibinfo {author} {\bibfnamefont {R.}~\bibnamefont
  {Orbach}},\ }\href {\doibase 10.1103/PhysRevLett.52.867} {\bibfield
  {journal} {\bibinfo  {journal} {Phys. Rev. Lett.}\ }\textbf {\bibinfo
  {volume} {52}},\ \bibinfo {pages} {867} (\bibinfo {year} {1984})}\BibitemShut
  {NoStop}%
\bibitem [{\citenamefont {Lois}\ \emph {et~al.}(2009)\citenamefont {Lois},
  \citenamefont {Blawzdziewicz},\ and\ \citenamefont {O'Hern}}]{Lois2009}%
  \BibitemOpen
  \bibfield  {author} {\bibinfo {author} {\bibfnamefont {G.}~\bibnamefont
  {Lois}}, \bibinfo {author} {\bibfnamefont {J.}~\bibnamefont {Blawzdziewicz}},
  \ and\ \bibinfo {author} {\bibfnamefont {C.~S.}\ \bibnamefont {O'Hern}},\
  }\href {\doibase 10.1103/PhysRevLett.102.015702} {\bibfield  {journal}
  {\bibinfo  {journal} {Phys. Rev. Lett.}\ }\textbf {\bibinfo {volume} {102}},\
  \bibinfo {pages} {015702} (\bibinfo {year} {2009})}\BibitemShut {NoStop}%
\bibitem [{\citenamefont {Ngai}\ \emph {et~al.}(1984)\citenamefont {Ngai},
  \citenamefont {Rajagopal},\ and\ \citenamefont {Huang}}]{Ngai1984}%
  \BibitemOpen
  \bibfield  {author} {\bibinfo {author} {\bibfnamefont {K.~L.}\ \bibnamefont
  {Ngai}}, \bibinfo {author} {\bibfnamefont {A.~K.}\ \bibnamefont {Rajagopal}},
  \ and\ \bibinfo {author} {\bibfnamefont {C.~Y.}\ \bibnamefont {Huang}},\
  }\href {\doibase http://dx.doi.org/10.1063/1.333452} {\bibfield  {journal}
  {\bibinfo  {journal} {Journal of Applied Physics}\ }\textbf {\bibinfo
  {volume} {55}},\ \bibinfo {pages} {1714} (\bibinfo {year}
  {1984})}\BibitemShut {NoStop}%
\bibitem [{\citenamefont {Gopal}(1966)}]{Gopal1966}%
  \BibitemOpen
  \bibfield  {author} {\bibinfo {author} {\bibfnamefont {E.}~\bibnamefont
  {Gopal}},\ }\href {http://books.google.ca/books?id=jAhRAAAAMAAJ} {\emph
  {\bibinfo {title} {Specific Heats at Low Temperatures}}},\ International
  cryogenics monograph series\ (\bibinfo  {publisher} {Plenum Press},\ \bibinfo
  {year} {1966})\BibitemShut {NoStop}%
\bibitem [{\citenamefont {Goetsch}\ \emph {et~al.}(2012)\citenamefont
  {Goetsch}, \citenamefont {Anand}, \citenamefont {Pandey},\ and\ \citenamefont
  {Johnston}}]{Goetsch2012}%
  \BibitemOpen
  \bibfield  {author} {\bibinfo {author} {\bibfnamefont {R.~J.}\ \bibnamefont
  {Goetsch}}, \bibinfo {author} {\bibfnamefont {V.~K.}\ \bibnamefont {Anand}},
  \bibinfo {author} {\bibfnamefont {A.}~\bibnamefont {Pandey}}, \ and\ \bibinfo
  {author} {\bibfnamefont {D.~C.}\ \bibnamefont {Johnston}},\ }\href {\doibase
  10.1103/PhysRevB.85.054517} {\bibfield  {journal} {\bibinfo  {journal} {Phys.
  Rev. B}\ }\textbf {\bibinfo {volume} {85}},\ \bibinfo {pages} {054517}
  (\bibinfo {year} {2012})}\BibitemShut {NoStop}%
\bibitem [{\citenamefont {Malinowski}\ \emph {et~al.}(2011)\citenamefont
  {Malinowski}, \citenamefont {Bezusyy}, \citenamefont {Minikayev},
  \citenamefont {Dziawa}, \citenamefont {Syryanyy},\ and\ \citenamefont
  {Sawicki}}]{Malinowski2011}%
  \BibitemOpen
  \bibfield  {author} {\bibinfo {author} {\bibfnamefont {A.}~\bibnamefont
  {Malinowski}}, \bibinfo {author} {\bibfnamefont {V.~L.}\ \bibnamefont
  {Bezusyy}}, \bibinfo {author} {\bibfnamefont {R.}~\bibnamefont {Minikayev}},
  \bibinfo {author} {\bibfnamefont {P.}~\bibnamefont {Dziawa}}, \bibinfo
  {author} {\bibfnamefont {Y.}~\bibnamefont {Syryanyy}}, \ and\ \bibinfo
  {author} {\bibfnamefont {M.}~\bibnamefont {Sawicki}},\ }\href {\doibase
  10.1103/PhysRevB.84.024409} {\bibfield  {journal} {\bibinfo  {journal} {Phys.
  Rev. B}\ }\textbf {\bibinfo {volume} {84}},\ \bibinfo {pages} {024409}
  (\bibinfo {year} {2011})}\BibitemShut {NoStop}%
\bibitem [{\citenamefont {Dormann}\ \emph {et~al.}(1988)\citenamefont
  {Dormann}, \citenamefont {Bessais},\ and\ \citenamefont
  {Fiorani}}]{Dormann1988}%
  \BibitemOpen
  \bibfield  {author} {\bibinfo {author} {\bibfnamefont {J.~L.}\ \bibnamefont
  {Dormann}}, \bibinfo {author} {\bibfnamefont {L.}~\bibnamefont {Bessais}}, \
  and\ \bibinfo {author} {\bibfnamefont {D.}~\bibnamefont {Fiorani}},\ }\href
  {http://stacks.iop.org/0022-3719/21/i=10/a=019} {\bibfield  {journal}
  {\bibinfo  {journal} {Journal of Physics C: Solid State Physics}\ }\textbf
  {\bibinfo {volume} {21}},\ \bibinfo {pages} {2015} (\bibinfo {year}
  {1988})}\BibitemShut {NoStop}%
\bibitem [{\citenamefont {Tholence}(1984)}]{Tholence1984}%
  \BibitemOpen
  \bibfield  {author} {\bibinfo {author} {\bibfnamefont {J.-L.}\ \bibnamefont
  {Tholence}},\ }\href {\doibase
  http://dx.doi.org/10.1016/0378-4363(84)90159-1} {\bibfield  {journal}
  {\bibinfo  {journal} {Physica B+C}\ }\textbf {\bibinfo {volume} {126}},\
  \bibinfo {pages} {157 } (\bibinfo {year} {1984})}\BibitemShut {NoStop}%
\bibitem [{\citenamefont {Souletie}\ and\ \citenamefont
  {Tholence}(1985)}]{Souletie1985}%
  \BibitemOpen
  \bibfield  {author} {\bibinfo {author} {\bibfnamefont {J.}~\bibnamefont
  {Souletie}}\ and\ \bibinfo {author} {\bibfnamefont {J.~L.}\ \bibnamefont
  {Tholence}},\ }\href {\doibase 10.1103/PhysRevB.32.516} {\bibfield  {journal}
  {\bibinfo  {journal} {Phys. Rev. B}\ }\textbf {\bibinfo {volume} {32}},\
  \bibinfo {pages} {516} (\bibinfo {year} {1985})}\BibitemShut {NoStop}%
\bibitem [{\citenamefont {Shtrikman}\ and\ \citenamefont
  {Wohlfarth}(1981)}]{Shtrikman1981}%
  \BibitemOpen
  \bibfield  {author} {\bibinfo {author} {\bibfnamefont {S.}~\bibnamefont
  {Shtrikman}}\ and\ \bibinfo {author} {\bibfnamefont {E.}~\bibnamefont
  {Wohlfarth}},\ }\href {\doibase
  http://dx.doi.org/10.1016/0375-9601(81)90441-2} {\bibfield  {journal}
  {\bibinfo  {journal} {Physics Letters A}\ }\textbf {\bibinfo {volume} {85}},\
  \bibinfo {pages} {467 } (\bibinfo {year} {1981})}\BibitemShut {NoStop}%
\bibitem [{\citenamefont {Vijayanandhini}\ \emph {et~al.}(2009)\citenamefont
  {Vijayanandhini}, \citenamefont {Simon}, \citenamefont {Pralong},
  \citenamefont {Caignaert},\ and\ \citenamefont {Raveau}}]{Vijay2009}%
  \BibitemOpen
  \bibfield  {author} {\bibinfo {author} {\bibfnamefont {K.}~\bibnamefont
  {Vijayanandhini}}, \bibinfo {author} {\bibfnamefont {C.}~\bibnamefont
  {Simon}}, \bibinfo {author} {\bibfnamefont {V.}~\bibnamefont {Pralong}},
  \bibinfo {author} {\bibfnamefont {V.}~\bibnamefont {Caignaert}}, \ and\
  \bibinfo {author} {\bibfnamefont {B.}~\bibnamefont {Raveau}},\ }\href
  {\doibase 10.1103/PhysRevB.79.224407} {\bibfield  {journal} {\bibinfo
  {journal} {Phys. Rev. B}\ }\textbf {\bibinfo {volume} {79}},\ \bibinfo
  {pages} {224407} (\bibinfo {year} {2009})}\BibitemShut {NoStop}%
\bibitem [{\citenamefont {Sow}\ and\ \citenamefont {Kumar}(2013)}]{Sow2013}%
  \BibitemOpen
  \bibfield  {author} {\bibinfo {author} {\bibfnamefont {C.}~\bibnamefont
  {Sow}}\ and\ \bibinfo {author} {\bibfnamefont {P.~S.~A.}\ \bibnamefont
  {Kumar}},\ }\href {http://stacks.iop.org/0953-8984/25/i=49/a=496001}
  {\bibfield  {journal} {\bibinfo  {journal} {Journal of Physics: Condensed
  Matter}\ }\textbf {\bibinfo {volume} {25}},\ \bibinfo {pages} {496001}
  (\bibinfo {year} {2013})}\BibitemShut {NoStop}%
\bibitem [{\citenamefont {Hoshi}\ \emph {et~al.}(2007)\citenamefont {Hoshi},
  \citenamefont {Katori}, \citenamefont {Kosaka},\ and\ \citenamefont
  {Takagi}}]{Hoshi2007}%
  \BibitemOpen
  \bibfield  {author} {\bibinfo {author} {\bibfnamefont {T.}~\bibnamefont
  {Hoshi}}, \bibinfo {author} {\bibfnamefont {H.~A.}\ \bibnamefont {Katori}},
  \bibinfo {author} {\bibfnamefont {M.}~\bibnamefont {Kosaka}}, \ and\ \bibinfo
  {author} {\bibfnamefont {H.}~\bibnamefont {Takagi}},\ }\href {\doibase
  http://dx.doi.org/10.1016/j.jmmm.2006.10.845} {\bibfield  {journal} {\bibinfo
   {journal} {Journal of Magnetism and Magnetic Materials}\ }\textbf {\bibinfo
  {volume} {310}},\ \bibinfo {pages} {e448 } (\bibinfo {year} {2007})},\
  \bibinfo {note} {proceedings of the 17th International Conference on
  Magnetism The International Conference on Magnetism}\BibitemShut {NoStop}%
\bibitem [{\citenamefont {Reuvekamp}\ \emph {et~al.}(2014)\citenamefont
  {Reuvekamp}, \citenamefont {Kremer}, \citenamefont {K\"ohler},\ and\
  \citenamefont {Bussmann-Holder}}]{Reuvekamp2014}%
  \BibitemOpen
  \bibfield  {author} {\bibinfo {author} {\bibfnamefont {P.~G.}\ \bibnamefont
  {Reuvekamp}}, \bibinfo {author} {\bibfnamefont {R.~K.}\ \bibnamefont
  {Kremer}}, \bibinfo {author} {\bibfnamefont {J.}~\bibnamefont {K\"ohler}}, \
  and\ \bibinfo {author} {\bibfnamefont {A.}~\bibnamefont {Bussmann-Holder}},\
  }\href {\doibase 10.1103/PhysRevB.90.094420} {\bibfield  {journal} {\bibinfo
  {journal} {Phys. Rev. B}\ }\textbf {\bibinfo {volume} {90}},\ \bibinfo
  {pages} {094420} (\bibinfo {year} {2014})}\BibitemShut {NoStop}%
\bibitem [{\citenamefont {Diaz}\ \emph {et~al.}(2004)\citenamefont {Diaz},
  \citenamefont {de~Brion}, \citenamefont {Holzapfel}, \citenamefont
  {Chouteau},\ and\ \citenamefont {Strobel}}]{Diaz2004}%
  \BibitemOpen
  \bibfield  {author} {\bibinfo {author} {\bibfnamefont {S.}~\bibnamefont
  {Diaz}}, \bibinfo {author} {\bibfnamefont {S.}~\bibnamefont {de~Brion}},
  \bibinfo {author} {\bibfnamefont {M.}~\bibnamefont {Holzapfel}}, \bibinfo
  {author} {\bibfnamefont {G.}~\bibnamefont {Chouteau}}, \ and\ \bibinfo
  {author} {\bibfnamefont {P.}~\bibnamefont {Strobel}},\ }\href {\doibase
  10.1016/j.physb.2004.01.038} {\bibfield  {journal} {\bibinfo  {journal}
  {Physica B: Condensed Matter}\ }\textbf {\bibinfo {volume} {346-347}},\
  \bibinfo {pages} {146} (\bibinfo {year} {2004})}\BibitemShut {NoStop}%
\bibitem [{\citenamefont {Hofmeister}(1991)}]{Hofmeister1991}%
  \BibitemOpen
  \bibfield  {author} {\bibinfo {author} {\bibfnamefont {A.~M.}\ \bibnamefont
  {Hofmeister}},\ }\href {\doibase 10.1029/91JB01381} {\bibfield  {journal}
  {\bibinfo  {journal} {Journal of Geophysical Research: Solid Earth}\ }\textbf
  {\bibinfo {volume} {96}},\ \bibinfo {pages} {16181} (\bibinfo {year}
  {1991})}\BibitemShut {NoStop}%
\bibitem [{\citenamefont {van~der Marck}(1997)}]{Marck1997}%
  \BibitemOpen
  \bibfield  {author} {\bibinfo {author} {\bibfnamefont {S.~C.}\ \bibnamefont
  {van~der Marck}},\ }\href {\doibase 10.1103/PhysRevE.55.1514} {\bibfield
  {journal} {\bibinfo  {journal} {Phys. Rev. E}\ }\textbf {\bibinfo {volume}
  {55}},\ \bibinfo {pages} {1514} (\bibinfo {year} {1997})}\BibitemShut
  {NoStop}%
\bibitem [{\citenamefont {Fiorani}\ \emph {et~al.}(1979)\citenamefont
  {Fiorani}, \citenamefont {Gastaldi}, \citenamefont {Lapiccirrella},
  \citenamefont {Viticoli},\ and\ \citenamefont {Tomassini}}]{Fiorani1979}%
  \BibitemOpen
  \bibfield  {author} {\bibinfo {author} {\bibfnamefont {D.}~\bibnamefont
  {Fiorani}}, \bibinfo {author} {\bibfnamefont {L.}~\bibnamefont {Gastaldi}},
  \bibinfo {author} {\bibfnamefont {A.}~\bibnamefont {Lapiccirrella}}, \bibinfo
  {author} {\bibfnamefont {S.}~\bibnamefont {Viticoli}}, \ and\ \bibinfo
  {author} {\bibfnamefont {N.}~\bibnamefont {Tomassini}},\ }\href {\doibase
  http://dx.doi.org/10.1016/0038-1098(79)90765-8} {\bibfield  {journal}
  {\bibinfo  {journal} {Solid State Communications}\ }\textbf {\bibinfo
  {volume} {32}},\ \bibinfo {pages} {831 } (\bibinfo {year}
  {1979})}\BibitemShut {NoStop}%
\bibitem [{\citenamefont {Fritsch}\ \emph {et~al.}(2012)\citenamefont
  {Fritsch}, \citenamefont {Yamani}, \citenamefont {Chang}, \citenamefont
  {Qiu}, \citenamefont {Copley}, \citenamefont {Ramazanoglu}, \citenamefont
  {Dabkowska},\ and\ \citenamefont {Gaulin}}]{Fritsch2012}%
  \BibitemOpen
  \bibfield  {author} {\bibinfo {author} {\bibfnamefont {K.}~\bibnamefont
  {Fritsch}}, \bibinfo {author} {\bibfnamefont {Z.}~\bibnamefont {Yamani}},
  \bibinfo {author} {\bibfnamefont {S.}~\bibnamefont {Chang}}, \bibinfo
  {author} {\bibfnamefont {Y.}~\bibnamefont {Qiu}}, \bibinfo {author}
  {\bibfnamefont {J.~R.~D.}\ \bibnamefont {Copley}}, \bibinfo {author}
  {\bibfnamefont {M.}~\bibnamefont {Ramazanoglu}}, \bibinfo {author}
  {\bibfnamefont {H.~A.}\ \bibnamefont {Dabkowska}}, \ and\ \bibinfo {author}
  {\bibfnamefont {B.~D.}\ \bibnamefont {Gaulin}},\ }\href {\doibase
  10.1103/PhysRevB.86.174421} {\bibfield  {journal} {\bibinfo  {journal} {Phys.
  Rev. B}\ }\textbf {\bibinfo {volume} {86}},\ \bibinfo {pages} {174421}
  (\bibinfo {year} {2012})}\BibitemShut {NoStop}%
\bibitem [{\citenamefont {Havlin}\ \emph {et~al.}(2014)\citenamefont {Havlin},
  \citenamefont {Kenett}, \citenamefont {Bashan}, \citenamefont {Gao},\ and\
  \citenamefont {Stanley}}]{Havlin2014}%
  \BibitemOpen
  \bibfield  {author} {\bibinfo {author} {\bibfnamefont {S.}~\bibnamefont
  {Havlin}}, \bibinfo {author} {\bibfnamefont {D.}~\bibnamefont {Kenett}},
  \bibinfo {author} {\bibfnamefont {A.}~\bibnamefont {Bashan}}, \bibinfo
  {author} {\bibfnamefont {J.}~\bibnamefont {Gao}}, \ and\ \bibinfo {author}
  {\bibfnamefont {H.}~\bibnamefont {Stanley}},\ }\href {\doibase
  10.1140/epjst/e2014-02251-6} {\bibfield  {journal} {\bibinfo  {journal} {The
  European Physical Journal Special Topics}\ }\textbf {\bibinfo {volume}
  {223}},\ \bibinfo {pages} {2087} (\bibinfo {year} {2014})}\BibitemShut
  {NoStop}%
\bibitem [{\citenamefont {Buldyrev}\ \emph {et~al.}(2010)\citenamefont
  {Buldyrev}, \citenamefont {Parshani}, \citenamefont {Paul}, \citenamefont
  {Stanley},\ and\ \citenamefont {Havlin}}]{Buldyrev2010}%
  \BibitemOpen
  \bibfield  {author} {\bibinfo {author} {\bibfnamefont {S.~V.}\ \bibnamefont
  {Buldyrev}}, \bibinfo {author} {\bibfnamefont {R.}~\bibnamefont {Parshani}},
  \bibinfo {author} {\bibfnamefont {G.}~\bibnamefont {Paul}}, \bibinfo {author}
  {\bibfnamefont {H.~E.}\ \bibnamefont {Stanley}}, \ and\ \bibinfo {author}
  {\bibfnamefont {S.}~\bibnamefont {Havlin}},\ }\href
  {http://dx.doi.org/10.1038/nature08932} {\bibfield  {journal} {\bibinfo
  {journal} {Nature}\ }\textbf {\bibinfo {volume} {464}},\ \bibinfo {pages}
  {1025} (\bibinfo {year} {2010})}\BibitemShut {NoStop}%
\bibitem [{\citenamefont {Bashan}\ \emph {et~al.}(2013)\citenamefont {Bashan},
  \citenamefont {Berezin}, \citenamefont {Buldyrev},\ and\ \citenamefont
  {Havlin}}]{Bashan2013}%
  \BibitemOpen
  \bibfield  {author} {\bibinfo {author} {\bibfnamefont {A.}~\bibnamefont
  {Bashan}}, \bibinfo {author} {\bibfnamefont {Y.}~\bibnamefont {Berezin}},
  \bibinfo {author} {\bibfnamefont {S.~V.}\ \bibnamefont {Buldyrev}}, \ and\
  \bibinfo {author} {\bibfnamefont {S.}~\bibnamefont {Havlin}},\ }\href
  {http://dx.doi.org/10.1038/nphys2727} {\bibfield  {journal} {\bibinfo
  {journal} {Nature Physics}\ }\textbf {\bibinfo {volume} {9}},\ \bibinfo
  {pages} {667} (\bibinfo {year} {2013})}\BibitemShut {NoStop}%
\bibitem [{\citenamefont {Stauffer}\ and\ \citenamefont
  {Aharony}(1994)}]{Stauffer1994}%
  \BibitemOpen
  \bibfield  {author} {\bibinfo {author} {\bibfnamefont {D.}~\bibnamefont
  {Stauffer}}\ and\ \bibinfo {author} {\bibfnamefont {A.}~\bibnamefont
  {Aharony}},\ }\href {https://books.google.ca/books?id=v66plleij5QC} {\emph
  {\bibinfo {title} {Introduction To Percolation Theory}}}\ (\bibinfo
  {publisher} {Taylor \& Francis},\ \bibinfo {year} {1994})\BibitemShut
  {NoStop}%
\bibitem [{\citenamefont {Wills}\ \emph {et~al.}(1998)\citenamefont {Wills},
  \citenamefont {Harrison}, \citenamefont {Mentink}, \citenamefont {Mason},\
  and\ \citenamefont {Tun}}]{Wills1998}%
  \BibitemOpen
  \bibfield  {author} {\bibinfo {author} {\bibfnamefont {A.~S.}\ \bibnamefont
  {Wills}}, \bibinfo {author} {\bibfnamefont {A.}~\bibnamefont {Harrison}},
  \bibinfo {author} {\bibfnamefont {S.~A.~M.}\ \bibnamefont {Mentink}},
  \bibinfo {author} {\bibfnamefont {T.~E.}\ \bibnamefont {Mason}}, \ and\
  \bibinfo {author} {\bibfnamefont {Z.}~\bibnamefont {Tun}},\ }\href
  {http://stacks.iop.org/0295-5075/42/i=3/a=325} {\bibfield  {journal}
  {\bibinfo  {journal} {EPL (Europhysics Letters)}\ }\textbf {\bibinfo {volume}
  {42}},\ \bibinfo {pages} {325} (\bibinfo {year} {1998})}\BibitemShut
  {NoStop}%
\bibitem [{\citenamefont {Ramirez}\ \emph {et~al.}(1990)\citenamefont
  {Ramirez}, \citenamefont {Espinosa},\ and\ \citenamefont
  {Cooper}}]{Ramirez1990}%
  \BibitemOpen
  \bibfield  {author} {\bibinfo {author} {\bibfnamefont {A.~P.}\ \bibnamefont
  {Ramirez}}, \bibinfo {author} {\bibfnamefont {G.~P.}\ \bibnamefont
  {Espinosa}}, \ and\ \bibinfo {author} {\bibfnamefont {A.~S.}\ \bibnamefont
  {Cooper}},\ }\href {\doibase 10.1103/PhysRevLett.64.2070} {\bibfield
  {journal} {\bibinfo  {journal} {Phys. Rev. Lett.}\ }\textbf {\bibinfo
  {volume} {64}},\ \bibinfo {pages} {2070} (\bibinfo {year}
  {1990})}\BibitemShut {NoStop}%
\bibitem [{\citenamefont {Ramirez}\ \emph {et~al.}(1992)\citenamefont
  {Ramirez}, \citenamefont {Espinosa},\ and\ \citenamefont
  {Cooper}}]{Ramirez1992}%
  \BibitemOpen
  \bibfield  {author} {\bibinfo {author} {\bibfnamefont {A.~P.}\ \bibnamefont
  {Ramirez}}, \bibinfo {author} {\bibfnamefont {G.~P.}\ \bibnamefont
  {Espinosa}}, \ and\ \bibinfo {author} {\bibfnamefont {A.~S.}\ \bibnamefont
  {Cooper}},\ }\href {\doibase 10.1103/PhysRevB.45.2505} {\bibfield  {journal}
  {\bibinfo  {journal} {Phys. Rev. B}\ }\textbf {\bibinfo {volume} {45}},\
  \bibinfo {pages} {2505} (\bibinfo {year} {1992})}\BibitemShut {NoStop}%
\bibitem [{\citenamefont {Mulder}\ \emph {et~al.}(1982)\citenamefont {Mulder},
  \citenamefont {van Duyneveldt},\ and\ \citenamefont {Mydosh}}]{Mulder1982}%
  \BibitemOpen
  \bibfield  {author} {\bibinfo {author} {\bibfnamefont {C.~A.~M.}\
  \bibnamefont {Mulder}}, \bibinfo {author} {\bibfnamefont {A.~J.}\
  \bibnamefont {van Duyneveldt}}, \ and\ \bibinfo {author} {\bibfnamefont
  {J.~A.}\ \bibnamefont {Mydosh}},\ }\href {\doibase 10.1103/PhysRevB.25.515}
  {\bibfield  {journal} {\bibinfo  {journal} {Phys. Rev. B}\ }\textbf {\bibinfo
  {volume} {25}},\ \bibinfo {pages} {515} (\bibinfo {year} {1982})}\BibitemShut
  {NoStop}%
\bibitem [{\citenamefont {Mulder}\ \emph {et~al.}(1981)\citenamefont {Mulder},
  \citenamefont {van Duyneveldt},\ and\ \citenamefont {Mydosh}}]{Mulder1981}%
  \BibitemOpen
  \bibfield  {author} {\bibinfo {author} {\bibfnamefont {C.~A.~M.}\
  \bibnamefont {Mulder}}, \bibinfo {author} {\bibfnamefont {A.~J.}\
  \bibnamefont {van Duyneveldt}}, \ and\ \bibinfo {author} {\bibfnamefont
  {J.~A.}\ \bibnamefont {Mydosh}},\ }\href {\doibase 10.1103/PhysRevB.23.1384}
  {\bibfield  {journal} {\bibinfo  {journal} {Phys. Rev. B}\ }\textbf {\bibinfo
  {volume} {23}},\ \bibinfo {pages} {1384} (\bibinfo {year}
  {1981})}\BibitemShut {NoStop}%
\bibitem [{\citenamefont {Stewart}\ \emph {et~al.}(2004)\citenamefont
  {Stewart}, \citenamefont {Ehlers}, \citenamefont {Wills}, \citenamefont
  {Bramwell},\ and\ \citenamefont {Gardner}}]{Stewart2004}%
  \BibitemOpen
  \bibfield  {author} {\bibinfo {author} {\bibfnamefont {J.~R.}\ \bibnamefont
  {Stewart}}, \bibinfo {author} {\bibfnamefont {G.}~\bibnamefont {Ehlers}},
  \bibinfo {author} {\bibfnamefont {A.~S.}\ \bibnamefont {Wills}}, \bibinfo
  {author} {\bibfnamefont {S.~T.}\ \bibnamefont {Bramwell}}, \ and\ \bibinfo
  {author} {\bibfnamefont {J.~S.}\ \bibnamefont {Gardner}},\ }\href
  {http://stacks.iop.org/0953-8984/16/i=28/a=L01} {\bibfield  {journal}
  {\bibinfo  {journal} {Journal of Physics: Condensed Matter}\ }\textbf
  {\bibinfo {volume} {16}},\ \bibinfo {pages} {L321} (\bibinfo {year}
  {2004})}\BibitemShut {NoStop}%
\end{thebibliography}%

\end{document}